\documentclass[aps,prd,twocolumn,superscriptaddress,nofootinbib,showpacs]{revtex4-1}
\usepackage[pass,letterpaper]{geometry}
\pdfoutput=1 
\usepackage[]{graphicx}
\usepackage[x11names]{xcolor}
\usepackage{slashed}
\usepackage{mathtools}
\usepackage{latexsym,amsmath,amsfonts,amssymb,bm}
\usepackage[pdfencoding=auto]{hyperref}
\usepackage{cancel}
\usepackage{booktabs}
\usepackage{verbatim}
\usepackage{hyperref}
\hypersetup{
  colorlinks,
  linkcolor={Brown4},
  citecolor={PaleGreen4},
  urlcolor={DodgerBlue2}
}

\def\dd{\textrm{d}}

\begin{document}

\title{Scalar-tensor theories at different scales: averaging the scalar sector}

\author{Jean-Philippe Uzan}
\email{uzan@iap.fr}
\affiliation{Institut d'Astrophysique de Paris, CNRS UMR 7095,
Sorbonne Universit\'es, 98 bis Bd Arago, 75014 Paris, France}
\affiliation{Center for Gravitational Physics and Quantum Information,
Yukawa Institute for Theoretical Physics, Kyoto University, 606-8502, Kyoto (Japan)}

\author{Hugo Lévy}
\email{hugo.levy@iap.fr}
\affiliation{Institut d'Astrophysique de Paris, CNRS UMR 7095,
Sorbonne Universit\'es, 98 bis Bd Arago, 75014 Paris, France}

\date{\today}  
\begin{abstract}
This article investigates the averaging of a scalar degree of freedom that couples universally to matter. It quantifies the approximation of smoothing the matter distribution before solving the Klein--Gordon equation. In the case of Yukawa theories, which enjoy a linear Klein--Gordon equation, the averaging commutes with the field equation as one might expect. While all small-scale distributions of matter lead to field distributions with the same mean, the latter can have different energy densities and pressures when the Compton wavelength of the field is smaller than the smoothing scale. In the non-linear case, such as chameleon theories, this study quantifies the error made by averaging the matter distribution before solving the Klein--Gordon equation. While field fluctuations can become arbitrarily large when the matter source is screened, the commutativity property of linear theories is recovered in the unscreened regime. Implications for cosmology --- and in particular the equation of state in extended quintessence models ---  and for laboratory experiments in low density medium are discussed. This analysis, although based on a simplifying description, sheds light on the effects of the small-scale distribution of matter as well as on the care required to define their equation of state and their cosmological signature.
\end{abstract}

\maketitle

\section{Averaging gravity theories}

In General Relativity, the Einstein field equations relate the spacetime geometry, {\em i.e.} the metric tensor $g_{\mu\nu}$, to the matter distribution described by its stress-energy tensor $T_{\mu\nu}$. As
emphasized by Ellis~\cite{Ellis:1971pg}, this involves, implicitly or explicitly, a smoothing scale in order to define the matter density, pressure and other properties.

General relativity is a non-linear theory. If the Einstein equations are supposed to hold on a local scale, say $s=1$ so that the spacetime geometry is solution of  $G_{\mu\nu}[g_{\alpha\beta}^{s=1}] =
\kappa T_{\mu\nu}^{s=1}$, its geometry smoothed on a larger scale, $s=2$ say, shall be obtained as some average of the small-scale geometry. However, one usually smooths the matter distribution before solving the Einstein equation to determine the large-scale geometry as $G_{\mu\nu}[g_{\alpha\beta}^{s=2}] = \kappa \langle T_{\mu\nu}^{s=1}\rangle_{s=2}$. This is the case in particular in cosmology since, based on the cosmological principle, one infers the most general form the energy-momentum tensor should have on large-scales, (with in particular the form of a perfect fluid imposed by the symmetries), and then solves the Einstein equations to obtain a large-scale metric of the Friedmann--Lema\^{\i}tre (FL) form. Indeed $G_{\mu\nu}[g_{\alpha\beta}^{s=2}] \not=G_{\mu \nu} [\langle g_{\alpha\beta}^{s=1}\rangle_{s=2}]$.

This leads to the so-called {\it averaging problem}, {\em i.e.} how the coarse-graining of matter, that hides small-scale effects, modifies the Einstein equations and how the geometries of the same spacetimes seen at different scales are related. It comes with a series questions among which {\it the backreaction problem}, {\em i.e.} the understanding of how local gravitational inhomogeneities may affect the cosmological dynamics and ultimately to {\it the fitting problem}~\cite{Ellis:1984bqf,Ellis:1987zz} which questions the determination of the best fit FL geometry. We refer to Refs.~\cite{Ellis:2005uz,Clarkson:2011zq} for reviews on the various approaches to these deep questions. Many roads to attack the averaging problem have been developed in the past years. It includes the Buchert formalism~\cite{Buchert:1999er,Buchert:2001sa},  well-adapted to dust filled universes, the Zalaletdinov formalism~\cite{Zalaletdinov:1996aj,Zalaletdinov:2008ts} that is an attempt to average the complete set of Cartan structure equations as well as approaches based on the deformation of the spatial metric of initial data sets along its Ricci flow~\cite{Paranjape:2008ai,Buchert:2002ht} or on the renormalisation group approach for coarse grained geometries~\cite{Calzetta:1999zr}; see also Refs.~\cite{Carfora:1995fj,Futamase:1996fk,Boersma:1997yt,Stoeger:1999ig} for earlier proposals.

Without having to go that far, it has been shown that the clumpiness of matter may bias the determination of the cosmological parameters~\cite{Fleury:2013sna,Fleury:2013uqa,Ben-Dayan:2012ccq} due to its impact on light propagation which is related to the so-called Ricci--Weyl problem, {\em i.e.} to the puzzling question to explain to which extent can the propagation of a light beam, which mostly occurs in regions dominated by Weyl curvature, can be addressed using the FL geometry, whose curvature is exclusively Ricci. It has been extensively  discussed in the context of cosmology~\cite{RWrefs,Clarkson:2011br} and recently been resolved thanks to the formalism proposed in Refs.~\cite{Fleury:2015rwa,Fleury:2017owg,Fleury:2018cro}.

There is a particular case in which the backreaction problem can be fully addressed: scalar gravity theories, {\em i.e.} Nordstr\"om gravity. Even though we know they do not offer a proper description of gravity and are ruled out by Solar system experiments, they enjoy the properties of having a single scalar degree of freedom that is non-minimally coupled to matter. The field equations then reduce to a  single Klein--Gordon (KG) equations with a potential and a coupling to matter. Both are functions of the scalar field so that the corresponding KG equation can be non-linear.  The goal of our study is to investigate the effect of averaging on the scalar sector of scalar tensor theories. For a generic matter distribution $\rho({\bf x})$ and prescribed boundary conditions, the KG equation determines the scalar field distribution $\phi({\bf x})$. Now, imagine one is interested in the system coarse grained on a scale $s$. The question that arises is how well is the response $\phi[\bar\rho_s]$ of the KG equation with the  coarse-grained matter source $\bar\rho_s$ a good approximation of the coarse-grained value of the true response of the KG equation, {\em i.e.}, $\bar\phi_s[\rho]$. Indeed, any non-linearities in the KG equation results in the non commutation of the averages so that one expects $\bar\phi_s[\rho]\not=\phi[\bar\rho_s]$. One shall however want to quantify the amplitude of the error that is made by averaging the matter sector before solving the KG equation instead of averaging the true microscopic field distribution. This is the first step, since then one shall also evaluate the same effect on the fluid variables, density and pressure, that enter the Einstein equations. Indeed since they are at least quadratic in the scalar field, one expects them to be highly sensitive to the matter distribution, even though the scalar field keeps the same large-scale average.

We shall thus consider two main models. First, the usual Yukawa interaction, that is a massive field with linear coupling to matter, enjoys a linear KG equation so that one expects the coarse-graining to commute with the field equation. But indeed, the small-scale distribution of matter will impact the large-scale field energy density and pressure even though it does not affect the smoothed field distribution. Second, the chameleon model is a non-linear theory that possesses a screening mechanism. Let us remind that minimally coupled light scalar field has been shown to remain smooth even in the presence of dark matter clumps~\cite{Ferreira:1997au} but it can cluster on scales larger than the horizon. Models with almost vanishing sound speed~\cite{Creminelli:2008wc} have been argued to lead to clustering of the quintessence field in the nonlinear regime~\cite{Creminelli:2009mu}. The situation drastically changes in the context of extended quintessence~\cite{Uzan:1999ch}, in particular if a screening mechanism based on non-linearities is at work.

This question is important at least in two contexts. First, the study of atomic transitions in the presence of a chameleon field~\cite{Levy:2024vyd} identified the difficulty of predicting the scalar field behavior in non-homogeneous media, which is yet crucial {\em e.g.} for properly modeling vacuum chamber experiments or for setting boundary conditions in outer space. Hence the interpretation of experiments trying to tackle down a chameleon field in low density medium may be affected by the fact that the fluid approximation is not good enough. 

Second, in standard cosmology, matter on large scales is described by a continuous fluid while it is clear that it is composed of galaxies on small scales. The analysis of scalar-tensor theories usually assumes that the scalar field is smooth on cosmological scales. Most studies of the cosmological dynamics therefore take the mean matter density as the source term. For linear scalar field sectors, it raises the question of the computation of the energy density and pressure of the scalar field that enter the Friedmann equations, which is a typical backreaction problem. For non-linear scalar field sectors, it has been shown on the particular case of the chameleon~\cite{Brax:2004qh,Brax:2004px}, that, since the matter density is redshifts as $a^{-3}$, the minimum of the potential is increases with time, leading to the conclusion~\cite{Brax:2004qh} that  ``the mass of small fluctuations about the minimum satisfies $m\gg H$. Hence, the characteristic response time of the chameleon is much shorter than the time over which the potential evolves, and the evolution is therefore adiabatic. In other words, if the chameleon starts at the minimum of its effective potential, it then remains at the minimum as the latter evolves in time."  This is a property of the homogeneous field solution of the KG equation with smooth matter density source and not of the coarse-grained of the true scalar field distribution.  Both can however differ substantially, depending, as we shall demonstrate, on whether the system at stake is screened or not. While the averaging problem in chameleon theories was considered in Refs.~\cite{Mota:2006fz,Mota:2006ed} --- in which the author analytically derive a macroscopic Compton wavelength for the scalar field inside a screened body that is itself made of individual particles --- the original analytical approximations and numerics presented in our work will go beyond and draw new consequences. In this article, we say that the matter source is screened whenever the scalar field reaches the value that minimizes its effective potential within this source, so that the notion can also be applied to the Yukawa theory.

To evaluate the impact of the smoothing of the matter distribution in scalar-tensor theories, we build on the numerical tools developed in Refs.~\cite{Levy:2022xni,Levy:2024mut,Levy:2023tps} to solve the KG equation in non-linear scalar-tensor gravity theories such as the chameleon~\cite{Khoury:2003rn,Khoury:2003aq} and symmetron~\cite{Hinterbichler:2010es} models. Following the formalism proposed in Refs.~\cite{Sanghai:2015wia,Sanghai:2016ucv,Sanghai:2017yyn,Sikora:2018imr,Briddon:2024ftz} we  adopt a very simplified view by describing the matter distribution as a lattice the size of which can be identified to the smoothing scale. Each cell contains a single spherical matter source of constant density. Hence one could quantify the effect of the properties of the sub-grid mass distribution while the mean density remains identical. The case of a Yukawa field was studied in Ref.~\cite{Fleury:2016tsz} to evaluate the gravitational binding energy and its role in cosmology.

This article will thus investigate the effect of the small-scale matter distribution of the field distribution, its large-scale averaged value and its fluid properties. In the case of the chameleon model, it will highlight the importance of determining whether the matter source is screened or not. In some regime, we identify orders of magnitude effects compared to their values derived after the matter source is smoothed. Even though, we do not solve the full field dynamics and restrict to static situation, this draws the attention on the standard  hypothesis to use a homogeneous scalar field evolving in a homogenous isotropic FL spacetime when investigating the late time acceleration of the universe if the quintessence field is coupled to matter or simply dark matter~\cite{Pitrou:2023swx,Uzan:2023dsk}. These effects can actually become important at late time while negligible in the early universe. A particular attention will be devoted to the equation of state of dark energy, whose value may significantly be affected by the matter distribution. This could be an interesting avenue to discuss the Hubble tension. In the case of experiments in low density media, this study questions the value of the field in the space between the particles constituting the gas, in particular when the systems exhibit large field fluctuations. Hence in the first situation, one is mostly interested in the energy density while in the second situation the field distribution is the relevant quantity.\\

The article is organized as follows. Section~\label{sec2}  describes  the scalar sector of a scalar-tensor theory focusing on its KG equation and then the geometrical set-up for our numerical computations. Section~\ref{sec3}
considers the case of a linear KG equation, namely a Yukawa theory to show that, as expected, the averaging procedure does commute with the field equation. Still it stresses that many field distributions with same average have different macroscopic properties (energy density, pressure, equation of state). Then, Section~\ref{sec4} consider the case of a non-linear theory, namely a chameleon model. We gather the derivation of equations and the description of the numerical scheme in Appendices.

\section{Averaging the scalar sector}\label{sec2}

We consider a general scalar-tensor theory, as described in Appendix~\ref{appA}. It depends on two free functions of the scalar field $\phi$, the potential $V$ and the universal coupling function to matter $A$. We work in natural units, for which ${c = \hbar = 1}$ such that $\phi$ has dimension of a mass. In the Einstein frame, the KG equation for a static configuration takes the form
\begin{equation}\label{e.kg1}
\Delta\phi = V'(\phi) +  \rho \alpha(\phi)\equiv V'_{\rm eff},
\end{equation}
with $\alpha\equiv \dd\ln A/\dd\phi$ and the effective potential defined by
\begin{equation}
 V_{\rm eff} \equiv V(\phi)+ \frac{\beta\rho}{M_{\rm p}}\phi,
\end{equation}
with $M_{\rm p}\equiv1/\sqrt{8\pi G}$.

\subsection{Rescaling the Klein--Gordon equation}

We restrict our analysis to models in which
\begin{equation}\label{e.paraV}
V(\phi) = C \phi^p, \qquad
\ln A = \frac{\beta}{M_{\rm P} }\phi\,.
\end{equation}
The effective potential has a minimum in
\begin{equation}\label{e.phimin}
\phi^{p-1}_{\min} (\rho) = -\hbox{sgn}(p)\frac{\beta\rho}{|p| C M_{\rm P}},
\end{equation}
that depends on the local matter density. At the minimum of its potential, the field has a mass
\begin{align}
\left.\frac{\dd^2 V_{\rm eff}}{\dd\phi^2}\right\vert_{\phi_{\rm min}}\equiv m^2 (\rho) &= p(p-1)C\phi^{p-2}_{\min} (\rho)
\end{align}
and we define the Compton wavelength of the field as ${\lambda\equiv m^{-1}}$. This parameterization contains the two cases of a Yukawa fifth force and a chameleon model, respectively for
$$
(C,p)_{\rm Y}= \left(\frac{1}{2}m^2,2\right)
\quad\hbox{and}\quad
(C,p)_{\rm C}= \left(\Lambda^{4+n},-n\right)\,.
$$

\begin{figure}
\centering
\includegraphics[width=0.45\textwidth]{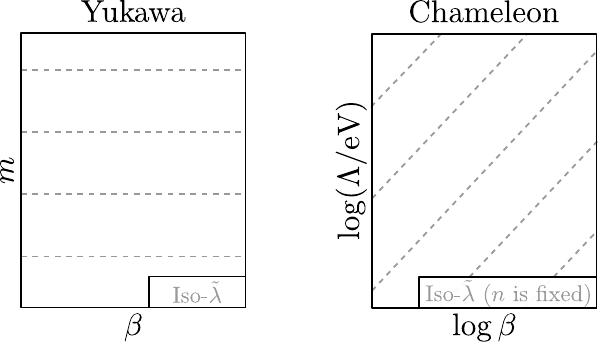}
\caption{Parameter spaces of the Yukawa and chameleon models. For each of them, the gray dashed lines correspond to iso-$\tilde\lambda$, the only parameter that enters the reduced Klein--Gordon equation~(\ref{e.KGred}).}
\label{fig:iso-alpha-models}
\end{figure}

In order to simplify the numerical integration, let us introduce a length scale $L_0$, the density and field scales, $\rho_0$ and $\phi_0$, all left unspecified for now. Then, we use the new space coordinates $\tilde{\bf x}\equiv {\bf x}/L_0$ and splits the density and scalar field as $\rho({\bf x}) =\rho_0 \tilde \rho(\tilde{\bf x})$ and $\phi({\bf x}) =\phi_0 \tilde \phi(\tilde{\bf x})$ with $(\tilde{\bf x},\tilde\rho,\tilde\phi)$ dimensionless quantities. Since Eq.~(\ref{e.kg1}) rewrites as
$$
\left[\frac{\phi_0}{L_0^2}\frac{M_{\rm P} }{\beta\rho_0}\right] \tilde\Delta \tilde\phi = \left[ \phi_0^{p-1} \frac{p c M_{\rm P} }{\beta\rho_0}\right] \tilde\phi^{p-1} + \tilde\rho\, ,
$$
it is convenient to choose $\phi_0$ so that the bracket on the r.h.s. reduces to $\rm{sgn}$$(p)$, which from Eq.~(\ref{e.phimin}) means 
$$
\left[-\phi_0^{p-1} /\phi_{\rm min}^{p-1}(\rho_0) \right]={\rm sgn}(p)\,.
$$
Note that $\phi_{\rm min}$ is not necessarily positive but has the sign of $-\hbox{sgn}(p)$. For the models considered here, Eq.~(\ref{e.phimin}) implies
\begin{equation}\label{e.phi0val}
\phi_{\min} (\rho) =  -\hbox{sgn}(p)  \phi_0
\end{equation}
with $\phi_0$ positive. It follows that the KG equation takes the generic form
\begin{equation}\label{e.KGred}
 \tilde\lambda^2 \tilde\Delta \tilde\phi =  {\rm sgn}(p) \tilde\phi^{p-1} + \tilde\rho
\end{equation}
with $\tilde\lambda^2=\tilde\lambda^2_{\mathrm Y}=1/m^2L_0^2=(\lambda/L_0)^2$ for the Yukawa model and $\tilde\lambda^2=\tilde\lambda^2_{\mathrm C}=(n+1)/m^2(\rho_0)L_0^2$ for the chameleon model. Figure~\ref{fig:iso-alpha-models} depicts the dependence of $\tilde\lambda$ on the model parameters.

Hence $\phi_0$ is a function of the model parameters $(C,p,\beta)$ and of the properties of the system at hand through the free parameter $\rho_0$, but it does not depend on $L_0$.  This non-dimensionalization procedure allows for a more efficient numerical exploration of these parameter spaces and lets us study a large variety of physical systems \textemdash \ in particular those of Table~\ref{tab1}.

\subsection{Geometrical set-up for the numerical integration}

Our goal is to evaluate the effect of the smoothing of the matter distribution on a typical scale $L_0$ on the field distribution as well as its energy and pressure. To that purpose, we define the average of any scalar quantity $X({\bf x})$ on the scale $s$ as
\begin{equation}
\bar X_s \equiv\langle X({\bf x}) \rangle_{s} = \frac{\int X({\bf y})W_{\! s}({\bf x}-{\bf y})\,\dd^3{\bf y}}{\int W_{\! s}({\bf x}-{\bf y})\,\dd^3{\bf y}}
\end{equation}
with $W_{\! s}$ a window function of typical width $s$.

\begin{figure}[t]
\centering
\includegraphics[width=\linewidth]{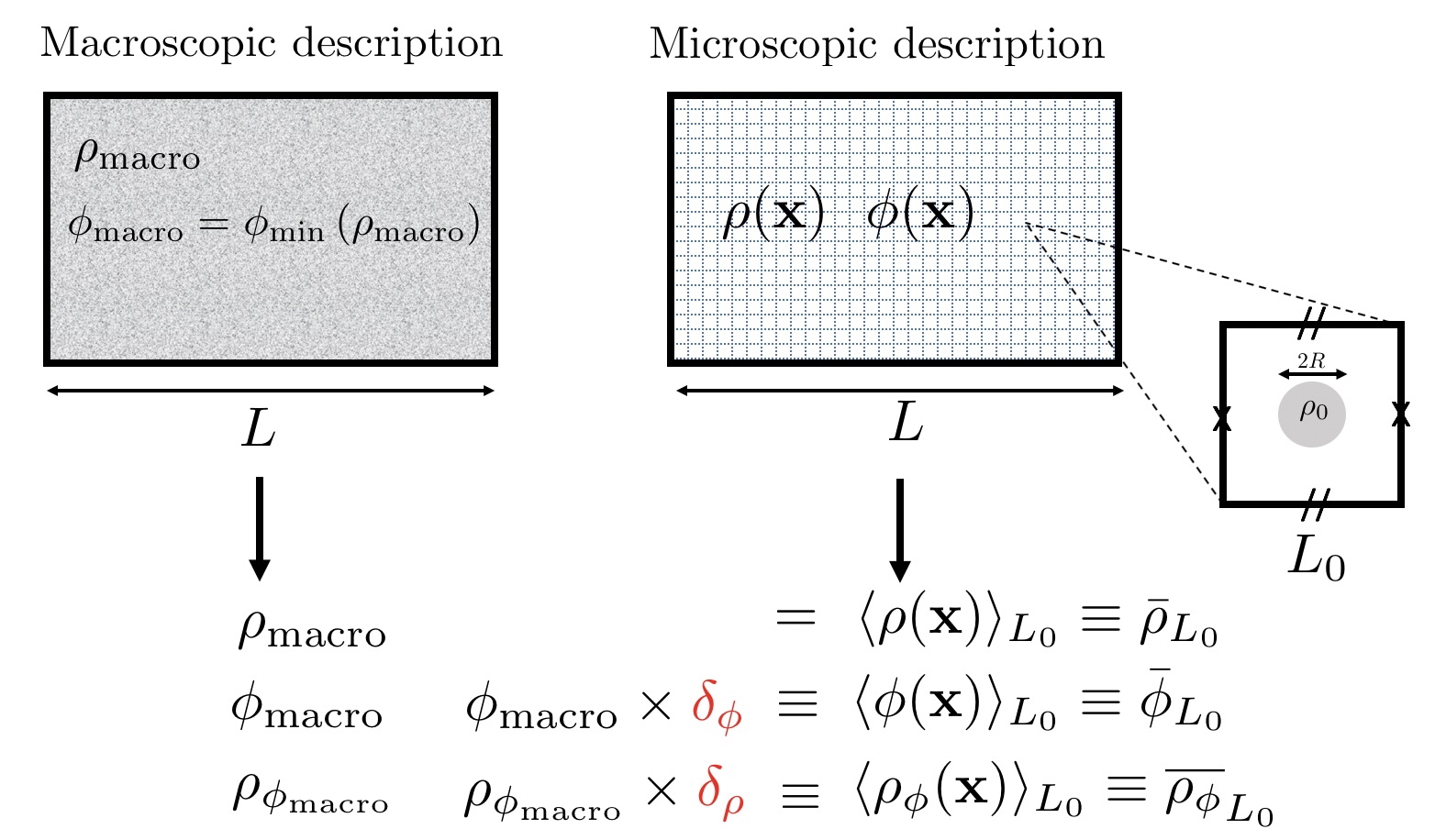}
\caption{Definition and relation of the microscopic and macroscopic descriptions related thanks to Eq.~(\ref{e.micromacro}). This also defines the two parameters $\delta_{\phi}$ and $\delta_{\rho}$  that give a measure of the non commutativity of the average with the computation of the local quantities; see Eq.~(\ref{e.defdelta}) with $\phi_{\rm macro}$ given by Eq.~(\ref{e.barphi}) and $\rho_{\phi_{\rm macro}}=V(\phi_{\rm macro})$.}
\label{fig:geometry}
\end{figure}

Now, we make the simplifying assumptions summarized on Fig.~\ref{fig:geometry} following the formalism by Refs.~\cite{Sanghai:2015wia,Sanghai:2016ucv,Sanghai:2017yyn,Sikora:2018imr,Briddon:2024ftz}. We consider a medium that is modeled  with two descriptions:
\begin{enumerate}
\item{\it Macroscopic.} On a large scales, the medium is homogeneous and continuous so that it can be described as a fluid with constant density $\rho_{\rm macro}$. For a static configuration the scalar field has thus the constant value 
\begin{equation}\label{e.barphi}
\phi_{\rm macro}\equiv \phi_{\min}(\rho_{\rm macro})\,,
\end{equation}
that is indeed not necessarily strictly positive; see Eq.~(\ref{e.phi0val}).
\item{\it Microscopic}. On the local scale, the fluid is described by spherical particles of radius $R$ with constant mass density $\rho_0$ in vacuum. For simplicity, we assume that these particles occupy the centers of a regular lattice of size $L_0$. For each representative elementary volume, the local matter distribution is thus given by $\rho(\mathbf{x}) \equiv 0$ if $\mathbf{x} \notin \mathcal{B}(R)$ and $\rho(\mathbf{x})=\rho_0$ otherwise, with $\mathcal{B}(R)$ the ball of radius $R$ centered on $\mathbf{0}$. The field distribution $\phi(\bf x)$ is then solution of the KG equation with this source term and periodic boundary conditions. To be more specific, we consider the domain of integration to $\Omega$ a cube of side length $L_0$ and boundary $\Gamma$. We also define $\tilde\Omega$ the cube of unit side length with boundary $\tilde\Gamma$ .
\end{enumerate}

\begin{table}[t]
\caption{Orders of magnitude of the physical features and derived parameters for three systems of interest.}
    \centering
    \begin{ruledtabular}
    \begin{tabular}{lccc}
        & Ambient air & Dilute gas & Universe \\\midrule
        Size of particles $R$ & $10^{-10} \, \mathrm{m}$ & $10^{-10} \, \mathrm{m}$ & $ 1\,\mathrm{kpc}$ \\
	Mean distance $L_0$ & $10^{-9} \, \mathrm{m}$ & $10^{-6} \, \mathrm{m}$ & $10^{3} \, \mathrm{kpc}$ \\
	Reduced size $\tilde{R}$ & $10^{-1}$ & $10^{-4}$ & $10^{-3}$ \\
	Local density $\rho_0$ & $10^3 \, \mathrm{kg/m^3}$ & $ 10^3 \, \mathrm{kg/m^3}$ & $ 10^{-18} \, \mathrm{kg/m^3}$ \\
	Mean density $\rho_{\mathrm{macro}}$ & $1 \, \mathrm{kg/m^3}$ & $10^{-10} \, \mathrm{kg/m^3}$ & $10^{-26} \, \mathrm{kg/m^3}$ \\
    \end{tabular}
    \end{ruledtabular}
\label{tab1}
\end{table}

The descriptions on these two scales are related by the fact that we shall impose ${\rho_{\rm macro} = \langle \rho({\bf x}) \rangle_{L_0}\equiv \bar \rho_{L_0}}$, hence for homogeneous distribution on ${\cal B}$
$$
\rho_{\rm macro} \hbox{Vol}(\Omega) = \rho_0  \hbox{Vol}({\cal B}),
$$
which for a sphere imposes
\begin{equation}\label{e.micromacro}
\rho_{\rm macro} = \frac{4\pi}{3}\left(\frac{R}{L_0}\right)^3\rho_0\equiv \frac{4\pi}{3}{\tilde R}^3\rho_0 \, .
\end{equation}
This expresses the fact that the fluid mass density $\rho_{\rm macro}$ is related to the mass density and size of the particles that compose it and their mean separation. 

Hence, the microscopic description of matter fixes $\rho_0$ and $R$ while the macroscopic description fixes $L_0$. Then, $\phi_0$ is algebraically determined by Eq.~(\ref{e.phi0val}) while $\phi_{\rm macro}$ is given by Eq.~(\ref{e.barphi}). The typical values of the physical  parameters for the systems of interest are gathered in Table~\ref{tab1}.\\

 In order to discuss the effect of the coarse graining, we shall compare the averages of the field distribution and energy density to the ones of the large-scale homogeneous description with $\rho_{\rm macro}$. We thus define the two quantities $\delta_{\phi}$ and $\delta_{\rho}$ by
\begin{equation}\label{e.defdelta}
\bar\rho_{L_0}=\rho_{\rm macro},\,\,\,\,\,
\bar\phi_{L_0} = \delta_{\phi}\phi_{\rm macro}, \,\,\,\,
{\overline{\rho_{\phi}}}_{L_0}= \delta_{\rho} \rho_{\phi_{\rm macro}}
\end{equation}
with $\phi_{\rm macro}$ given by Eq.~(\ref{e.barphi}) and  $ \rho_{\phi_{\rm macro}}=V(\phi_{\rm macro})$ the energy density of the smooth field distribution. We use the similar quantity $\delta_P$ for the isotropic pressure.

In a quadratic scalar field theory such as the Yukawa theory, i.e. with a linear KG equation, one expects $\delta_{\phi}=1$ while $\delta_{\rho}\not=1$ and $\delta_{P}\not=1$ whatever the theory since the energy density is a quadratic function of the field. Indeed non-linear scalar field theories such as the chameleon model  will lead to $\delta_{\phi}\not=1$. The goal of the two following sections is to estimate the magnitude of these three quantities.

\begin{figure*}[t]
\centering
\includegraphics[width=\textwidth]{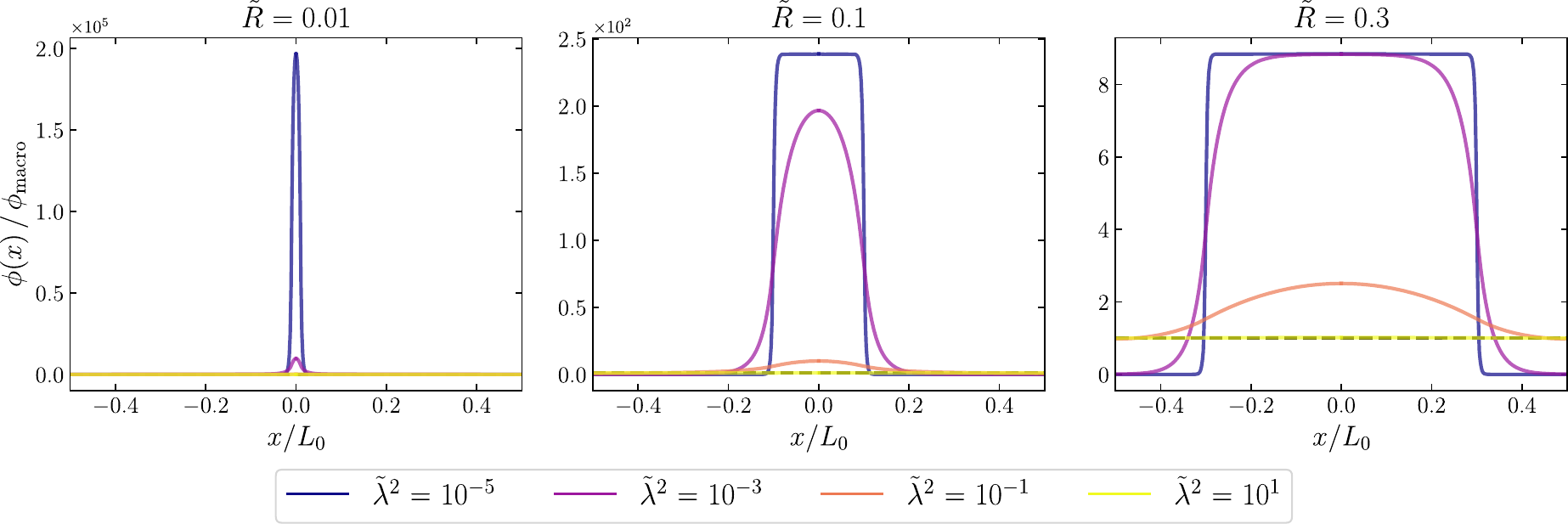}
\caption{Yukawa field profile $\phi({\bf x})$ normalized by $\phi_{\rm macro}$ evaluated along one of the principal axis of the cube $x$. The radius $R$ and microscopic density $\rho_0$ are varied such that $R^3\rho_0$ remains constant, {\em i.e.} all microscopic geometric configurations share the same value of $\rho_{\rm macro}$, {\em i.e.} the same macroscopic fluid description. From left to right, $\tilde{R}=R/L_0$ takes the values  $0.01$, $0.1$ and $0.3$. The model parameters $(\lambda , \, \beta)$ are varied such that $\tilde\lambda^2=(\lambda/L_0)^2$  spans  from $10^{-5}$ to $10$ (colored solid lines). It is easily checked that whatever the Yukawa parameters, all these geometric configurations have $\delta_{\phi}=1$ (black dashed line) with the same $\phi_{\rm macro}$. Note that both $\phi(x)$ and $\phi_{\mathrm{macro}}$ are negative, which is why $\phi / \phi_{\mathrm{macro}} >0$.}
\label{fig3}
\end{figure*}

\section{Example of a quadratic theory: the Yukawa field}\label{sec3}

Let us start by the study of a theory in which the KG equation remains linear. This is the case of a Yukawa field, that is a massive field linearly and universally coupled to matter.

\subsection{Distribution of the scalar field and its average}\label{subsec3a}

First, before even having to solve the KG equation, it is clear that one expects from linearity that the averaging procedure shall commute with the equation. Indeed, from the divergence theorem, the l.h.s. of Eq.~(\ref{e.KGred}) gives
$$
\int_{\tilde \Omega}\tilde\Delta\tilde\phi \, \dd^3\tilde{\bf x}= \int_{\tilde \Gamma} \tilde{\boldsymbol{\nabla}} \tilde\phi \cdot {\bf n} \, \dd^2\tilde\Gamma = 0,
$$
the last equality following from the periodic boundary conditions [see Eq. (\ref{eqn:epbc})]. It follows that 
$$
\int_{\tilde \Omega}\tilde\phi\, \dd^3\tilde{\bf x}= - \int_{\tilde \Omega}\tilde\rho\, \dd^3\tilde{\bf x}\,,
$$
{\it i.e.}, $\langle\tilde\phi \rangle_{1} = -\langle \tilde\rho \rangle_{1}$, a relation that interestingly does not depend on the parameter $\tilde\lambda^2$. The average of the field distribution is thus given by $\bar\phi_{L_0}\equiv \langle\phi \rangle_{L_0} =$ $\phi_0 \langle\tilde\phi \rangle_{1}=$ $-\phi_0 \langle \tilde\rho \rangle_{1}=-\phi_0\rho_{\rm macro}/\rho_0$, so that, from Eqs.~(\ref{e.phimin}) and~(\ref{e.barphi}),
\begin{equation}
\bar\phi_{L_0}=\phi_{\rm macro} \quad \Longleftrightarrow \quad\delta_{\phi}=1
\end{equation}
as anticipated.\\

\begin{figure*}[t]
\centering
\includegraphics[width=\linewidth]{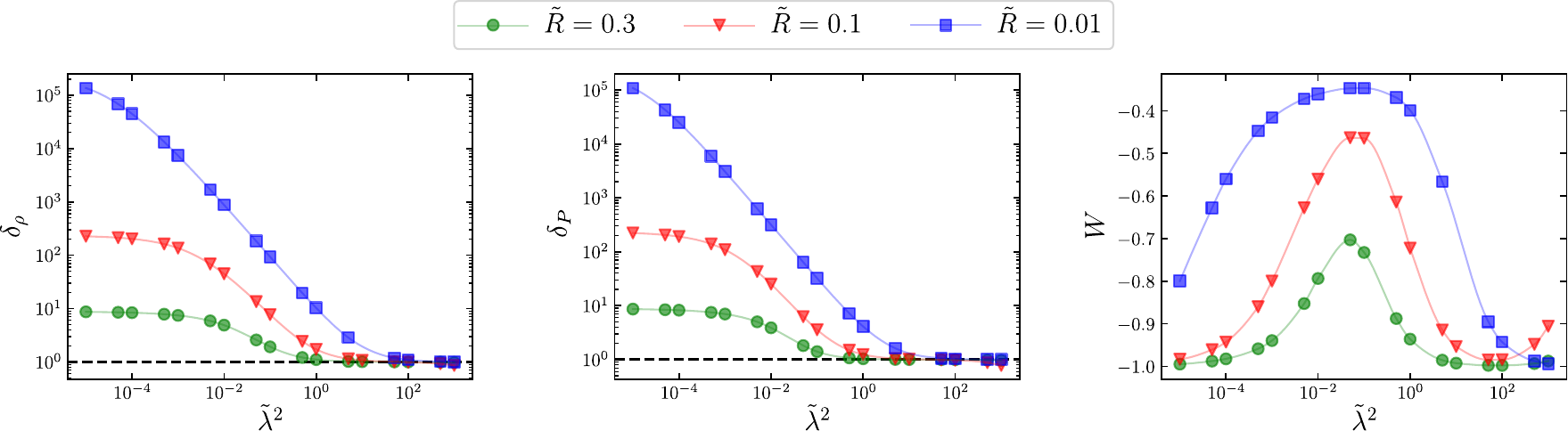}
\caption{Dependency of $\delta_\rho$ and $\delta_P$ as a function of the $\tilde\lambda=\lambda/L_0$. The effect of the averaging increases if the clumpiness of matter, {\em i.e.} when $\tilde R$ decreases. The parameters are varied as in Fig.~\ref{fig3} in such a way that all models have the same averaged value of the field, $\phi_{\rm macro}$. The profiles are obtained from Eqs.~(\ref{e.Bdrho1}) and~(\ref{e.BP1}). The last panel presents the effective equation of states, $W=-\delta_p/\delta_\rho$, which shows a significant deviation from $w=-1$ for the homogeneous scalar field. The interpolating solid lines connecting the data points are shown only to provide a better view of trends. The analytic approximation~(\ref{e.approx}) gives the values $2.4\times 10^5$, $2.4\times10^2$ and $8.8$ respectively for $\tilde R= (0.01,0.1,0.3)$ for the plateau of $\delta_\rho$ in the limit $ \lambda \ll 1$, perfectly fitting with the numerics.}
\label{fig:mean-rhophi_Yukawa}
\end{figure*}

To be more quantitative and compare to our numerical integration of the KG equation, let us remind that in the Yukawa theory, the linear differential operator ${(\Delta - 1/\lambda^2)}$ has a free-space Green function
$$
{\cal G}({\bf x},{\bf x}') = -\frac{1}{4\pi} \frac{\hbox{e}^{-|{\bf x}-{\bf x}'|/\lambda}}{|{\bf x}-{\bf x}'|}\,.
$$
Hence the field generated by a point particle source, i.e. with $\rho({\bf x})= M \delta_{\rm D}({\bf x})$ is
\begin{equation}
\phi_{\rm p.p.}({\bf x}) = -\frac{\beta M}{4\pi M_{\rm P}} \frac{\hbox{e}^{-r/\lambda}}{r}
\,.
\end{equation}
As detailed in Appendix~\ref{appC}, the field distribution generated by a sphere of radius $R$ and constant density $\rho_0$ is
\begin{equation}\label{e.sphere1}
\phi(r) =-\frac{1}{4\pi}\frac{\beta M}{M_{\rm P}}  \left\lbrace
\begin{tabular}{l}
$\Phi(R/\lambda) \frac{\hbox{e}^{-r/\lambda}}{r} $\\[8pt]
 $3\frac{1-\hbox{e}^{-R/\lambda}\frac{R+\lambda}{r}\sinh(r/\lambda) }{(R/\lambda)^3}\frac{1}{\lambda} $
 \end{tabular}
\right.,
\end{equation}
respectively oustside and inside the sphere, with ${M=4\pi \rho_0 R^3/3}$ and
\begin{equation}\label{e.PPHI}
\Phi(x) = 3(x \cosh x - \sinh x)/x^3,
\end{equation}
It is easily checked that this expression is indeed smooth in $r=R$ and that the usual Newtonian expressions are obtained in the limit $\lambda\rightarrow+\infty$.

These expressions allow us to qualitatively discuss the effect of the mass distribution, {\em i.e.} the choice of $(R,\rho_0)$ while $M$ is kept identical. First, inside the sphere, Eq.~(\ref{e.sphere1}) can be rewritten as
$$
\phi(r) =\phi_{\rm min} \left[ 1-\hbox{e}^{-R/\lambda}   (R/\lambda+1)   \frac{\sinh(r/\lambda)}{r/\lambda} \right]\,
$$
showing that the field is driven exponentially close to $\phi_{\rm min}$ if $R/\lambda$ is large enough, {\em i.e.}, if the field’s Compton wavelength to be much smaller than $R$. Typically $|\phi(0)/\phi_{\rm min}-1|<1\%$ for $R/\lambda\gtrsim 6.6$. Then, outside the the sphere, the field relaxes to $0$ but its amplitude is rescaled by the factor $\Phi(R/\lambda)$ compared to a point particle with same mass. All these features are numerically illustrated on Fig.~\ref{fig3} even though they take into account the periodic boundary conditions necessary to represent the infinite lattice. All the field profiles depicted on Fig.~\ref{fig3} have different $(R,\rho_0)$ with same total mass, {\em i.e.}  same $\rho_{\rm macro}$. It is easily checked that while  the field distributions $\phi({\bf x})$ depend on both $(R,\rho_0)$ and the parameter $(\lambda,\beta)$, they all share the same mean value $\bar\phi_{L_0}$, i.e. $\delta_{\phi}=1$ for all models and all parameters enjoying the same $\bar\rho_{L_0}$.

\subsection{Distribution of the energy  and its average}

Since the  stress energy tensor~(\ref{e.Tphi}) is quadratic in the scalar field, all these field profiles sharing the same mean value (see Fig.~\ref{fig3})  have different energy densities. Let us evaluate how the coarse grained energy depend on the small-scale distribution of matter, {\em i.e.}, on the parameters $(R,\rho_0)$ with same $\rho_0 R^3$. The general expression~(\ref{e.Brho1}) of the mean energy density of the scalar field, $\overline{\rho_\phi}_{L_0}$ reduces to
$$
\overline{\rho_\phi}_{L_0}=-\frac{\beta}{2M_{\rm P}L_0^3}\int_{{\cal B}} \rho({\bf x})\phi({\bf x})\,\dd^3{\bf x}
$$
for a Yukawa model (since $p=2$). This expression has been interpreted as a binding energy in Ref.~\cite{Fleury:2016tsz}. For a sphere, if one approximates the field by the spherically symmetric solution~(\ref{e.sphere1}) which indeed neglects the periodic boundary conditions, one can get the approximation
$$
\delta_\rho \simeq -\left(\frac{\phi_0}{\phi_{\rm macro}}\right)^2 \int_{0}^{\tilde R} \tilde\phi(\tilde r) 4\pi \tilde r^2\, \dd \tilde r
$$
with $\tilde\phi(\tilde r)\simeq-\left[1-\hbox{e}^{-\tilde R/\tilde\lambda}(\tilde R/\tilde \lambda+1) \frac{\sinh(\tilde r/\tilde\lambda)}{\tilde r/\tilde\lambda} \right]$ and $\tilde\rho=1$ on ${\cal B}(R)$. The $\simeq$ symbol is used to recall we neglect the boundary conditions imposed by the large-scale structure. Hence,
\begin{eqnarray}
\delta_\rho &\simeq& 4\pi \left(\frac{\phi_0}{\phi_{\rm macro}}\right)^2\!\! \tilde\lambda^3 \!\!  \int_0^{\frac{\tilde R}{\tilde \lambda}}\!\! \!\! u^2\dd u
\left[1-\hbox{e}^{-\frac{\tilde R}{\tilde \lambda}}\left(\frac{\tilde R}{\tilde \lambda}+1\right) \frac{\sinh u}{u} \right] \nonumber \\
&\simeq& \frac{3}{4\pi\tilde R^3} \left[1- \hbox{e}^{-\frac{\tilde R}{\tilde \lambda}}\left(\frac{\tilde R}{\tilde \lambda}+1\right)\Phi(\tilde R/\tilde \lambda)\right]
\end{eqnarray}
It follows that in the limit of small $\tilde\lambda$,
\begin{equation}\label{e.approx}
\delta_\rho \sim \frac{3}{4\pi\tilde R^3} \left[1-\frac{3}{2}\frac{\tilde \lambda}{\tilde R}+{\cal O}(2) \right]\,,
\end{equation}
which explains that $\delta_\rho$ reaches a plateau with an amplitude that scales as $\tilde R^{-3}$, as obtained numerically on Fig.~\ref{fig:mean-rhophi_Yukawa}. At large $\tilde\lambda$, $\delta_\rho\simeq 1$, which can be explained by the fact that the field is less screened. The deviation from $\delta_\rho=1$ increases with the clumpiness of matter, {\em i.e.} when $\tilde R$ decreases which is indeed intuitive since when  $4\pi\tilde R^3/3\rightarrow 1$, $\rho_0\rightarrow\rho_{\rm macro}$ and the sphere fills the whole domain. This analytic approximation has been checked to give the exact value of the plateau of Fig.~\ref{fig:mean-rhophi_Yukawa} in the limit $\tilde\lambda \ll 1$.

\subsection{Discussion of the physical implications}

This analysis can be extended to compute the coarse grained isotropic pressure.  The results are gathered on Fig.~\ref{fig:mean-rhophi_Yukawa} that show a similar behavior for $\delta_P$. We also consider the equation of state of the scalar field. On the macroscopic scales $P_\phi(\phi_{\rm macro})=-\rho_\phi(\phi_{\rm macro})$ so that $w_{\rm macro}=-1$, while the equation of state of the coarse-grained distribution is
\begin{equation}
W \equiv  {\overline{\rho_\phi}}_{L_0} / {\overline{P_\phi}}_{L_0}    = - \frac{\delta_\rho}{\delta_P}\,.
\end{equation}

In order to interpret these numerical results, notice that for large $\tilde\lambda$ the matter source remains {\it unscreened}. The field is almost not affected by the matter on the scale $\tilde R$ so that it can be treated perturbatively. Hence one concludes that both $\delta_\rho$ and $\delta_P$ tend toward 1 and hence $W=-1$.  For small $\tilde\lambda$, the matter sphere are {\it screened} so that the field reaches its two minima in the whole sphere and in the inter-sphere space --- but in a small layer $\Delta R$. It follows that the contribution of $(\nabla\phi)^2$ in the field pressure~(\ref{eq.Pphidef}) becomes negligible since $\epsilon\equiv\Delta R/R\ll1$. Hence the expressions~(\ref{eq.rhophidef}-\ref{eq.Pphidef}) lead to the conclusion that $W=-1$, as confirmed by the numerical integrations on Fig.~\ref{fig:mean-rhophi_Yukawa}.

These results drive a series of remarks.
\begin{enumerate}
\item In the cosmological context, the energy density and pressure of the scalar field smoothed on cosmological scales shall enter the Friedmann equation. This effect has been investigated in Ref.~\cite{,Briddon:2024ftz} in their ``post-Newtonian approach to cosmological modeling'', with which we share the same description of the microscopic distribution of matter. They concluded that the Friedmann equation shall take he generic form
$$
H^2 = \frac{8\pi G}{3}\left[\Big\langle\rho({\bf x}, t)\Big\rangle + \Big\langle\frac{1}{2}\dot\phi({\bf x}, t) +V[\phi({\bf x}, t)] \Big\rangle\right]
$$
which includes the fluctuations of the field on $\dot\phi$ but neglects the $\nabla\phi$ contribution to the scalar field energy density~(\ref{eq.rhophidef}).  This is indeed a good approximation when the post-Newtonian formalism can be applied, {\em i.e.}, gravity is described by a metric and that it does not involve any characteristic scale~\cite{will,blanchet} on the size of the system studied. in the present case, this corresponds to the limit $\tilde\lambda\gg1$, a regime in which we fully recover the conclusions of Ref.~\cite{,Briddon:2024ftz}. This is indeed not the case otherwise, as shown on Fig.~\ref{fig:mean-rhophi_Yukawa}. The $\tilde\lambda\leq1$ regime is indeed much more subtle to model. If we boldly assume that the Friedmann equations take the form -- which indeed is probably not the case,
\begin{eqnarray}
H^2 &=& \frac{8\pi G}{3}\left(\overline\rho + {\overline{\rho_\phi}} \right)\nonumber\\
\frac{\ddot a}{a} &=&-\frac{4\pi G}{3}\left(\overline\rho+3\overline P + {\overline{\rho_\phi}} +3 {\overline{P_\phi}}\right)\nonumber
\end{eqnarray}
as long as it is described in the Einstein frame, which is indeed not the observational frame --- see point~2 below -- then the computation of the full energy density is required, with the gradient terms included.

As already emphasized ${\overline{\rho_\phi}}\not= \rho_\phi(\phi_{\rm macro})$ and Fig.~\ref{fig:mean-rhophi_Yukawa} let us expect that they can significantly differ. Indeed, in our numerics are restricted to a static configuration and we not solve the KG equation on a FL background spacetime. Still, we can speculate that ({\em 1}) in perturbation theory this is a lower order effect and the cosmological dynamics is well described by the usual approach relying on a FL solution as already concluded in  Ref.~\cite{,Briddon:2024ftz}; ({\rm 2}) in the regime $\tilde\lambda\gg1$, studied in  Ref.~\cite{,Briddon:2024ftz}, it is a good approximation to consider the smooth value of the field to infer the cosmological dynamics and ({\em 3}) in the regime $\tilde\lambda\ll1$ while, $\delta_\rho$ and $\delta_P$ become large, we expect the cosmological dynamics to significantly differ from the one inferred under the approximation of a smooth scalar field, even if, quite surprizingly, the equation of state seems not to be modified deep in this regime.

Our present analysis does not allow us to draw definite conclusions, but it signals that the cosmological conclusions for dark energy models coupled to matter shall be revisited with care, maybe by extending the analysis of Ref.~\cite{,Briddon:2024ftz} or our present work.

\item An additional point is worth emphasizing. So far, we have used the Einstein-frame description but observations shall be analyzed in the Jordan frame~\cite{Esposito-Farese:2000pbo,Uzan:2006mf}. It follows that
$$
a_{\rm JF} = A(\phi) a_{\rm EF}\,
$$
and
$$
\rho_{\rm JF} = A^4(\phi) \rho_{\rm EF}, \,
P_{\rm JF} = A^4(\phi) P_{\rm EF}
$$
respectively for the scale factor, energy density and pressure in both frames. Since the coupling $A(\phi)$ is not linear, it is then obvious that
$$
\langle A(\phi) \rangle a_{\rm EF}\not= A(\phi_{\rm macro})  a_{\rm EF}
$$
and similarly
$$
\langle \rho_{\rm JF} \rangle = \langle A^4(\phi)  \rho_{\rm EF}\rangle \not= A^4(\phi_{\rm macro})  \langle \rho_{\rm EF}\rangle \,.
$$
Indeed, these effects are usually taken into account at linear order in cosmological perturbations but can become large when the clumpiness of matter becomes important.

\item As seen on Fig.~\ref{fig:mean-rhophi_Yukawa}, the equation of state  remains larger than $-1$, $W\geq-1$, and is still driving an acceleration since $W<-1/3$. Hence, one shall expect the effect of the clustering to lead to a milder acceleration. Even in the case of a Yukawa model with a large mass compared to the Hubble parameter, so that the macroscopic dynamics leads to a pure cosmological constant since the field is stuck at the minimum of its potential, one expects $W>-1$. Indeed our analysis is restricted to a static distribution so that it has to be extended to derived the proper cosmological effects. But it shows the effect is present.

\item It follows that extra-care is needed for the interpretation of extended quintessence models {\em even if} the KG equation is linear and $\bar\phi=\phi_{\rm macro}$ since the clumpiness of matter will be more important at small redshifts and hence modify the late time prediction of the observables. This also the case in models~\cite{Pitrou:2023swx,Uzan:2023dsk} in which the scalar field coupled simply to a dark matter component to avoid variations of fundamental constants~\cite{Uzan:2024ded}.
\end{enumerate}

Hence, this example, even if not realistic and restricted to a static matter distribution, shows that the averaging of the scalar can bias the interpretation of scalar field model when coupled to matter.

\section{Example of a non-linear theory: the chameleon field}\label{sec4}

\begin{figure*}[t]
\centering
\includegraphics[width=\textwidth]{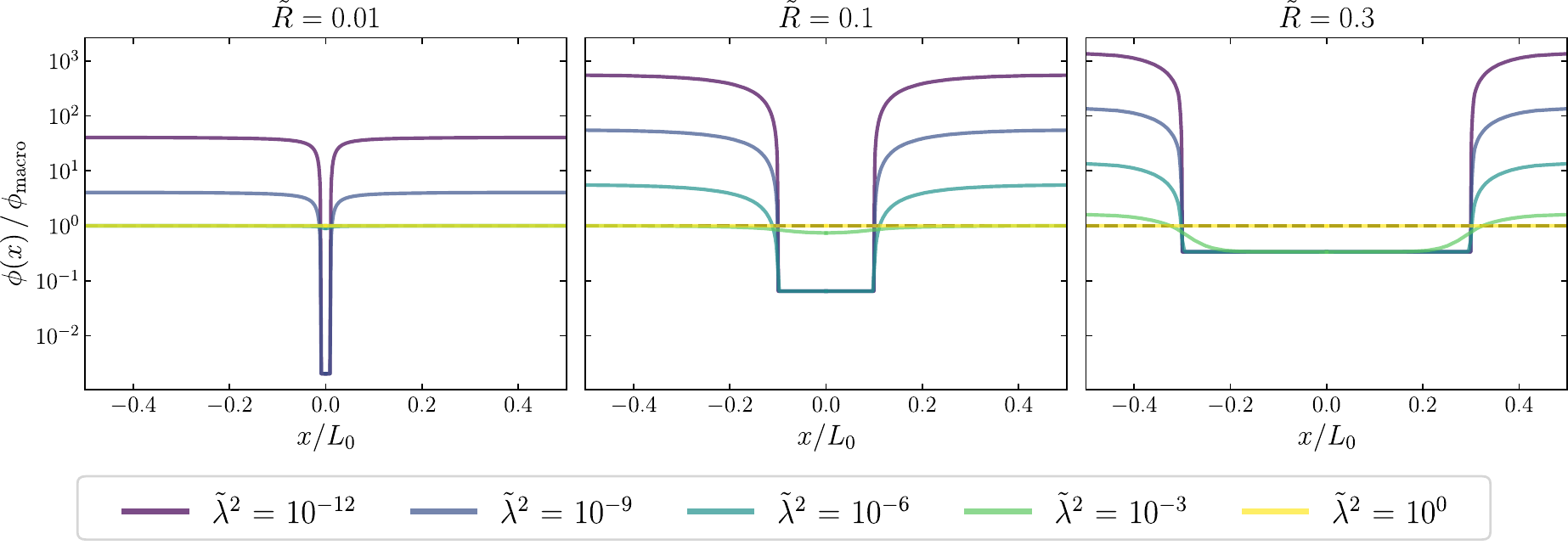}
\caption{Chameleon field profiles $\phi(x)$ normalized by $\phi_{\rm macro}$ evaluated along one of the principal axis of the cube $x$. The radius $R$ and microscopic density $\rho_0$ are varied such that $R^3\rho_0$ remains constant, {\em i.e.} all microscopic geometric configurations share the same value of $\rho_{\rm macro}$, i.e. the same macroscopic fluid description. From left to right, $\tilde R=R/L_0$ takes the values  $0.01$, $0.1$ and $0.3$. The model parameters $(\Lambda, \beta)$ are varied such that $\tilde{\lambda}$ spans from $10^{-10}$ to $1$ (colored solid lines). In the screened regime (e.g. for ${\tilde{\lambda}=10^{-10}}$), the average value of the field strongly depends on the microscopic matter distribution.}
\label{fig5}
\end{figure*}

\subsection{Distribution of the scalar field}

Let us now consider a non-linear KG equation as arises in chameleon models. Unfortunately, such a scenario cannot be fully addressed analytically. Still, repeating step by step the argument of \S~\ref{subsec3a}, one easily shows that
\begin{equation}
\langle \tilde\phi^{-(n+1)}\rangle_1 = \langle \tilde \rho \rangle_1\,,
\end{equation}
from which one concludes that, thanks to Eqs.~(\ref{e.phimin}) and~(\ref{e.phi0val})
\begin{equation}
\langle \phi^{-(n+1)}\rangle_{L_0} = \phi_{\rm macro}^{-(n+1)}.
\end{equation}
This shows that the field distribution with different microscopic geometries with constant $\bar\rho_{L_0}$ will still share some common averaged quantities, but they will indeed have different mean values of $\phi$ so that $\delta_\phi\not=1$. This generically implies, \textit{a fortiori}, that  the field energy density will have very different mean values.\\

\begin{figure*}[t]
\centering
\includegraphics[width=\textwidth]{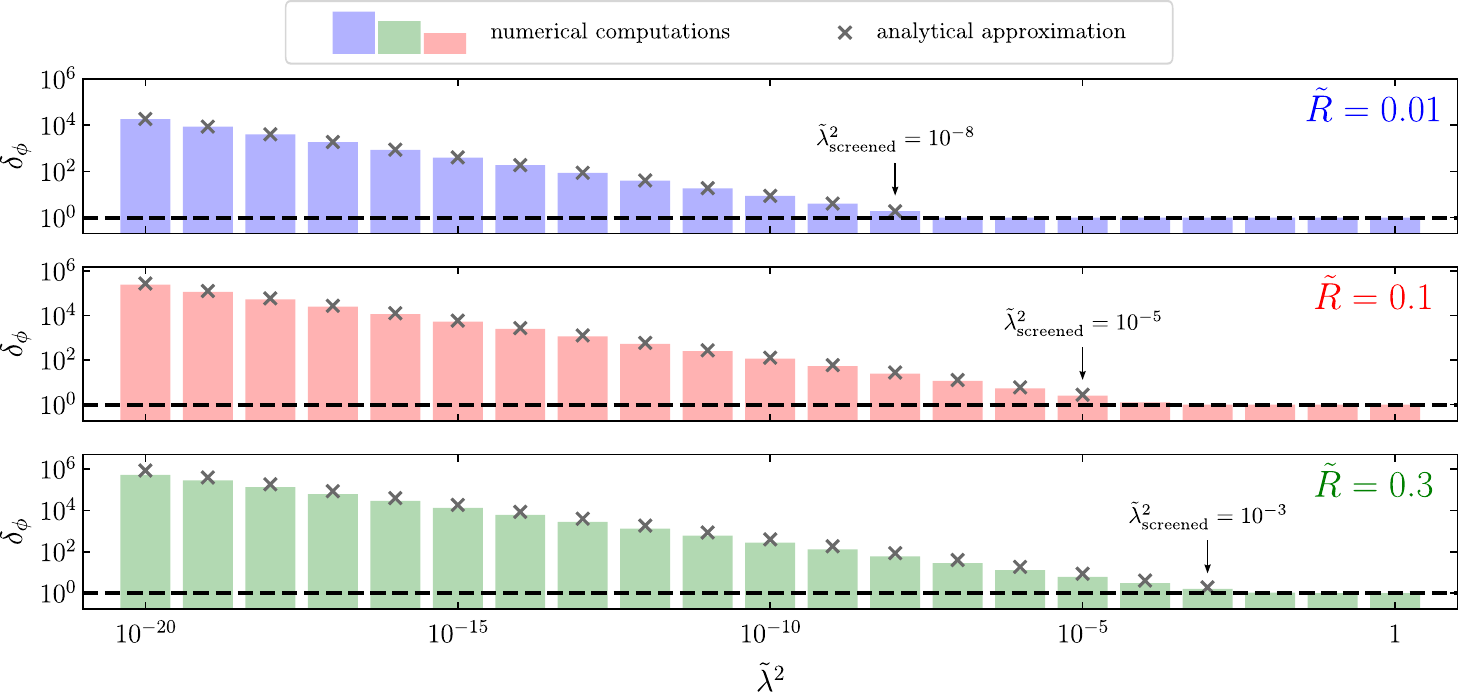}
\caption{Due to the non-linearities of the KG equations, chameleon models generically predict that $\langle \phi \rangle_{L_0} \not= \phi_{\rm macro}$  while $\langle \phi^{-(n+1)}\rangle_{L_0} = \phi_{\rm macro}^{-(n+1)}$, a relation that has indeed be checked to hold. The panels present the numerical computation of $\delta_\phi$, defined in Eq.~(\ref{e.defdelta}) as a function of $\tilde\lambda\equiv\lambda/L_0$ for $\tilde R= 0.01, 0.1$ and $0.3$ from top to bottom. We indicate the values of $\tilde\lambda$ that corresponds to the shift between the screened and unscreened regimes. The $\times$ symbols correspond the analytical approximation~(\ref{e.delta_phi_analytical}) valid in the screened regime. It is remarkable that in the unscreened regime $\delta_\phi=1$ despite the non-linearities.}
\label{fig6}
\end{figure*}

Figure~\ref{fig5} depicts several chameleon field profiles normalized by $\phi_{\mathrm{macro}}$ evaluated along one of the principal axis of the cube. Some of the comments made on the Yukawa field profiles (see Fig.~\ref{fig3}) still apply to the present case. In particular, the limits $\tilde{\lambda} \to 0^+$ and $\tilde{\lambda} \to + \infty$ have similar effects on the chameleon field dynamics: in the former limit, the field reaches the minimum of its effective potential inside the sphere; whereas in the latter limit, the field relaxes to the constant $\phi_{\mathrm{macro}}$ everywhere. The main difference with respect to the Yukawa model \textemdash \ as already emphasized \ \textemdash \ is that $\delta_{\phi} \neq 1$ in the so-called screened regime. This point is not necessarily obvious from Fig.~\ref{fig5}, which is why we compute $\delta_{\phi}$ for three distinct geometries and for a wide range of parameters $\tilde{\lambda}$ in Fig.~\ref{fig6}. Two comments are in order from theses numerical results:
\begin{enumerate}
\item When the system is {\em unscreened} on the smoothing scale $L_0$, which is typically the case when $\tilde{\lambda} \gtrsim 1$, one recovers $\delta_{\phi} = 1$ despite the non-linerarities.
\item In the {\em screened regime} however, {\em i.e.} for sufficiently small values of $\tilde{\lambda}$, $\delta_{\phi}$ departs from unity, reaching values larger than $10^5$.
\end{enumerate}

In the latter case, {\em i.e.} when the matter sphere is screened, we can even derive an analytical approximation of $\delta_{\phi}$. Following the steps detailed in Appendix~\ref{appD}, we get
\begin{equation}
\delta_{\phi} \simeq \left( \frac{4}{3} \pi \tilde{R}^3 \right)^{\! \frac{1}{n+1}} \left[ \frac{4}{3} \pi \tilde{R}^3 + \left( 1 - \frac{4}{3} \pi \tilde{R}^3 \right) \tilde{\phi}_{\mathrm{vac}} \right]
\label{e.delta_phi_analytical}
\end{equation}
with
\begin{equation}
\tilde{\phi}_{\mathrm{vac}} \simeq \left( \frac{4 \pi \tilde{R} \tilde{\lambda}^2}{1 - 4 \pi \tilde{R}^3/3} \right)^{\! -\frac{1}{n+2}} .
\label{e.phi_vac_analytical}
\end{equation}
This analytical approximation is depicted by gray crosses and compared against the numerical computations in Fig.~\ref{fig6}. The two approaches \textemdash \ analytical and numerical  \textemdash \ agree remarkably well. Consequently, it seems legitimate to make use of Eqs.~(\ref{e.delta_phi_analytical}) and (\ref{e.phi_vac_analytical}) to extrapolate values of $\delta_{\phi}$ for even smaller parameters $\tilde{\lambda}$. Moreover, Eq.~(\ref{e.delta_phi_analytical}) also turns out to be also useful for approximating $\tilde{\lambda}_{\mathrm{screened}}$, by checking when Eq.~(\ref{e.delta_phi_analytical}) produces $\delta_{\phi} \simeq 1$ (see gray crosses on Fig.~\ref{fig6}). Here, $\tilde{\lambda}_{\mathrm{screened}}$ refers to the specific value of $\tilde{\lambda}$ where the transition between the two regimes occurs.

Note that a similar analysis has been performed in Ref.~\cite{Briddon:2024ftz} and that our field profiles from Fig.~\ref{fig5} have the same generic form that those presented in their Fig.~5 even though we do not use the same boundary conditions. Still, deep in the screened regime, the two computation led to consistent conclusions since the parameter $\alpha$ in Eqs.~(48-29) of Ref.~\cite{Briddon:2024ftz} introduced in the averaged Friedmann equations (neglecting the gradient term in the energy density since it differs from 0 only in a thin shell)  coincides with our ${\rm Vol}(\tilde\Omega)=4 \pi \tilde{R}^3/3$. The present work gives however a precise, numerical and analytical, value on $\delta_\phi$, {\rm i.e.}, on how the small scale structure affects the mean value of the scalar field.

\subsection{Implications}

\begin{figure*}[t]
\centering
\includegraphics[width=\textwidth]{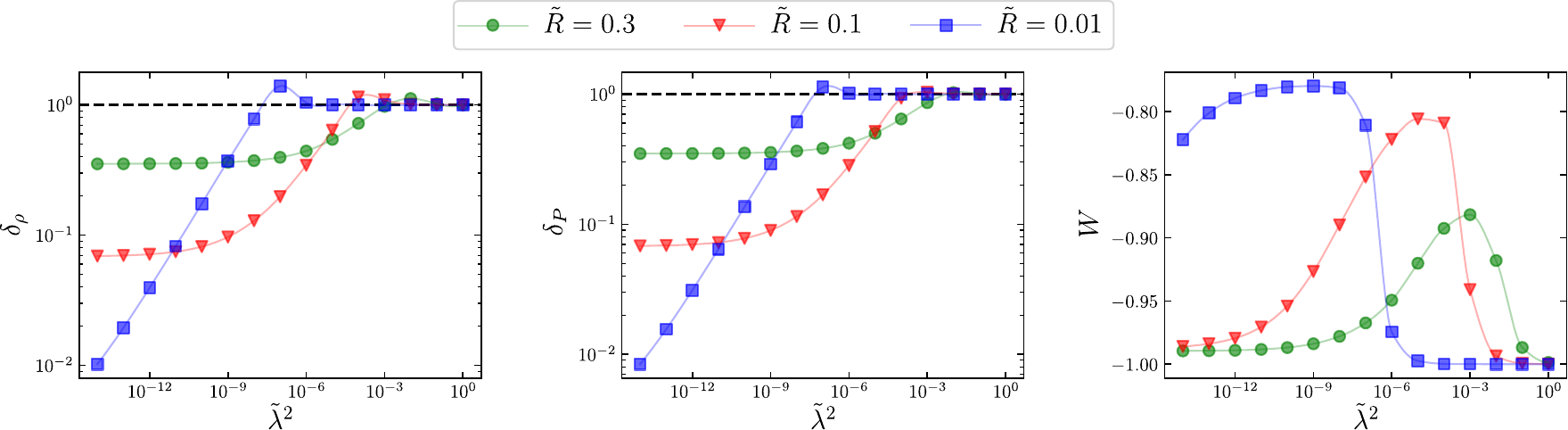}
\caption{Dependency of $\delta_\rho$ and $\delta_P$ as a function of the $\tilde\lambda=\lambda/L_0$. The effect of the averaging increases if the clumpiness of matter, {\em i.e.} when $\tilde R$ decreases. The parameters are varied as in Fig.~\ref{fig3} in such a way that all models have the same averaged value of the field, $\phi_{\rm macro}$. The profiles are obtained from Eqs.~(\ref{e.Bdrho1}) and~(\ref{e.BP1}). The last panel presents the effective equation of states, $W=-\delta_p/\delta_\rho$, which shows a significant deviation from $w=-1$ for the homogeneous scalar field. Its behaviors at large and small $\tilde\lambda$ are explained in the core of the text. The interpolating solid lines connecting the data points are shown only to provide a better view of trends.}
\label{fig7}
\end{figure*}

Figure~\ref{fig7} gives the consequences of the fluid quantities and the equation of state.  As  for the field distribution, the consequences depend strongly on the screening.

In the {\em unscreened} regime, both $\delta_\rho$ and $\delta_P$ tend to unity so that $W=-1$. Interestingly this is also the case in the {\em screened} regime, for the same reason as for the Yukawa case: the field has strong variations between the inside and outside of the sphere but it is almost constant everywhere but on a thin shell of relative width $\epsilon$. Hence its gradient is almost zero everywhere and the local equation of state is equal to $-1$ almost everywhere. 

\begin{figure*}[t]
\centering
\includegraphics[width=\textwidth]{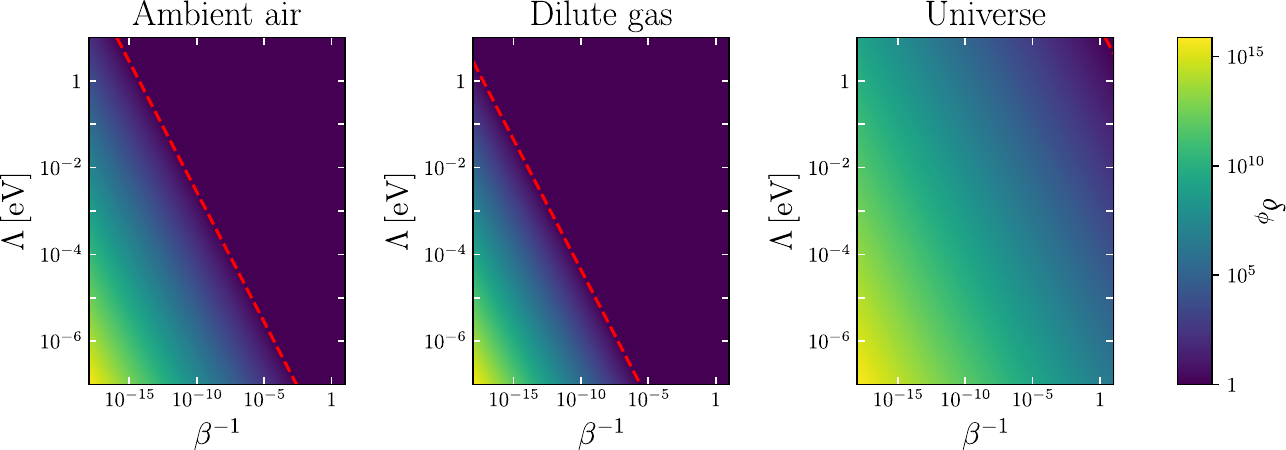}
\caption{$\delta_{\phi}$ in the chameleon parameter space for the three physical systems described in Table~\ref{tab1} [computed from the analytical approximation~(\ref{e.delta_phi_analytical})]. The red dashed line indicates the transition from the unscreened regime, for which $\delta_{\phi} = 1$, to the screened regime, for which one can have $\delta_{\phi} \gg 1$. The $x$-axis is $1/\beta$ to be in line with the exclusion plots found in the literature.}
\label{fig8}
\end{figure*}

Then, from Fig.~\ref{fig8}, we see that for  typical values found in Table~\ref{tab1}, galaxies are not screened, {\em i.e.} at late time the homogenous approximation used to draw cosmological consequences may not be a good. Hence, the field may  not be following the minimum of its effective potential $V_{\rm eff}$ with $\rho=\rho_{\rm macro}$, and the effects on the cosmological dynamics may be strongly modified. Still the equation of state is reduced but less sharply than for the Yukawa case.

\section{Discussion}\label{sec5}

This article has investigated the consequences of smoothing the matter distribution before solving the field equation of the scalar degree of freedom of a scalar-tensor theory of gravity. Its dynamics is described by a Klein--Gordon equation with a self interacting potential and a coupling to the energy density of the matter distribution. Two cases were considered. First a Yukawa theory, for which the KG equation remains linear, and then a chameleon model with a non-linear KG equation.

While the matter distribution is simplified to a static crystal structure containing a single homogeneous sphere, this toy model nonetheless allowed us to assess the effect of small-scale matter distribution on scalar field-dependent quantities (for a constant mean density value). As expected, the Yukawa theory yields that the average value of the field distribution matches the macroscopic value of the scalar field obtained from the mean energy density, {\em i.e.}, $\delta_\phi=1$. Still, the small-scale distribution affects the mean energy density and pressure, and thus eventually the equation of state of the effective fluid describing the scalar field on large-scales. For the non-linear theory it is demonstrated that this can be reached only if the matter source remains unscreened. In both cases, we have shown that $\delta_\rho= 1$ and $\delta_P=1$ in the unscreened regime so that effectively the average commutes with the equation of motion, even in the non-linear case. We have also shown that the equation of state is unaffected {\em deep} in either the unscreened or screened regimes.

Even if this toy model remains very simple, it highlights a series of questions both for cosmology and laboratory searches of a fifth force. As shown on Fig.~\ref{fig8}, for the various physical systems described in Table~\ref{tab1}, $\delta_{\phi} \gg 1$ in a large region of the chameleon parameter space. In this region, it is not valid to model such media as homogeneous fluids. As a consequence, it may be necessary to re-analyze several results in the literature.

First, our results have to be put into perspective with fifth force searches in the laboratory and in space; see {\em e.g.} Ref.~\cite{Burrage:2017qrf} for a review. In space, one usually assumes that the scalar field relaxes to $\phi_{\mathrm{min}} (\rho_{\mathrm{space}}) $ infinitely far away from the system of interest. This is a key assumption to derive the response of the detector as seen {\em e.g.} on the concrete example of the MICROSCOPE experiment~\cite{Pernot-Borras:2019gqs,Pernot-Borras:2020jev,Pernot-Borras:2021edr,Levy:2023tps}. Indeed, all studies aiming at constraining the chameleon model on Solar system scales, including the seminal Ref.~\cite{Khoury:2003rn}), do assume ${\phi \to \phi_{\infty}}$. Indeed, given the present discussion, in particular in the screened regime, the chosen asymptotic boundary condition affects the field's profile, together with its gradient which is proportional to the fifth force amplitude. Hence a great care shall be taken on the way to deal with accurate boundary conditions in a dilute media. Beyond fifth force searches, experiments relying on `potential effects' --- see Table~1 in Ref.~\cite{Levy:2024vyd} --- should also be re-analyzed more closely. Taking $\phi_{\mathrm{macro}}$ as the maximum value of the scalar field in outer space is only valid if $\delta_{\phi} = 1$ and one shall explicitly check that the fluid approximation for the surrounding media is good enough.

Then, concerning cosmology, the right-hand panel of Fig.~\ref{fig8} indicates that $\delta_{\phi} \gg 1$ almost everywhere in the parameter space. This has important consequences as already emphasized above. Most cosmological predictions are drawn using the assumption that the universe is well described by a FL geometry on all scales, using perturbation theory. Indeed, under that assumption, $\delta_\rho$ and $\delta_P$ would be second order in perturbations. At late time, when the clustering of matter becomes significant, our study indicate that (\textit{i}) $\delta_\phi$ may be large for non-linear theories in the screened regime and (\textit{ii}) $\delta_\rho$ and $\delta_P$ are usually different from unity, see Figs.~\ref{fig:mean-rhophi_Yukawa} and~\ref{fig7}, which can bias the evolution of the universe, even if the Friedman equations remain a good description of the cosmic dynamics. Interestingly we have shown that the small-scale distribution of matter affect the effective equation of state of the scalar field if it is non minimally coupled, which is the case in extended quintessence models. This also holds in the class of models in which it couples only to dark matter. To finish, we have pointed out that in any scalar-tensor theory, one should also keep in mind that the average does \textit{not} generally commute with the change of frame. Indeed we worked in Einstein frame for the sake of simplicity but observations have to be interpreted in the Jordan frame.

As a conclusion, this analysis questions the reliability of the cosmological and laboratory tests of scalar field models, in particular in the screened regime. A similar concern was pointed out by Ref.~\cite{Fleury:2016tsz}. Here, we have identified many situations in which the homogeneous fluid, and hence the homogeneous asymptotic value of the scalar field, does not hold. We have further made explicit computations --- both analytically and numerically --- of the order of magnitude of the error made by averaging the matter distribution before solving the KG equation. A concrete example at the atomic level, that actually drove the present analysis, is presented in Ref.~\cite{Levy:2024vyd}. Many developments, in particular to include the time evolution and the effect of an expanding spacetime as well as a more realistic matter distribution, will be addressed later.

\section*{Acknowledgements}

We thank Joel Berg\'e, George Ellis, Julien Larena and Cyril Pitrou for their insight and comments and Pierre Fleury for an in depth discussion to help us compare the results by Ref.~\cite{Briddon:2024ftz} to ours. HL thanks the Institute of astrophysics for hospitality during the end of this work.

\appendix
\section{Scalar-tensor theory}\label{appA}

Let us consider a general scalar-tensor theory, with the action
\begin{equation}
S = S_{\mathrm{EH}} + S_{\phi} + S_{\mathrm{mat}}
\end{equation}
with
\begin{subequations}
\begin{align}
S_{\mathrm{EH}} &= \frac{M_{\rm P}^2}{2} \int \mathrm{d}^4 x \sqrt{-g} R \, , \label{subeqn:einstein-hilbert-action} \\[3pt]
S_{\phi} &= - \int \mathrm{d}^4 x \sqrt{-g} \biggl[ \frac{1}{2} g^{\mu \nu} \partial_{\mu} \phi \, \partial_{\nu} \phi + V(\phi) \biggr] \, , \label{subeqn:scalar-field-action} \\[3pt]
S_{\mathrm{mat}} &= \int \mathrm{d}^4 x \sqrt{-\tilde{g}} \, \mathcal{L}_{\mathrm{mat}} \bigl( \tilde{g}_{\mu \nu} , \, \psi_{\mathrm{mat}}^{(i)} \bigr) \, . \label{subeqn:matter-action}
\end{align}
\label{eqn:action-parts}%
\end{subequations}
We assume that the metric tensor has ${(-, \, + , \, + , \, +)}$. Greek indices ${(\mu, \, \nu , \, \rho , \, \sigma , \, \text{etc.})}$ run from 0 to 3 while Latin indices ${(i , \, j , \, k , \, \text{etc.})}$ run from 1 to 3. This is indeed the standard Einstein\textendash Hilbert action with the Ricci scalar $R$ constructed from the Einstein-frame metric $g_{\mu \nu}$. Then, the second term~(\ref{subeqn:scalar-field-action}) is the action of the scalar field $\phi$ with a canonical kinetic term and potential $V$. Then, Eq.~(\ref{subeqn:matter-action}) denotes the matter action, where all matter fields $\psi_{\mathrm{mat}}^{(i)}$ are universally coupled to the Jordan-frame metric $\tilde{g}_{\mu \nu}$. The latter is chosen to be related to the Einstein-frame metric through a Weyl transformation, reading
\begin{equation}
\tilde{g}_{\mu \nu} = A^2 (\phi) g_{\mu \nu} \, ,
\label{eqn:weyl-transform}
\end{equation}
for some real conformal factor function $A$. It follows that the fields equations, obtained by varying the action with respect to $g^{\mu \nu}$ and $\phi$, reduce to
\begin{align}
G_{\mu \nu} \equiv R_{\mu \nu} - \frac{1}{2} R g_{\mu \nu} &= \frac{1}{M_{\rm P}^2} \bigl( T_{\mu \nu} + T_{\mu \nu}^{(\phi)} \bigr) \label{eqn:metric-eqn-ef} \\[3pt]
\Box \phi \equiv g^{\mu \nu} \nabla_{\!\mu} \! \nabla_{\! \nu} \phi &= \frac{\mathrm{d} V}{\mathrm{d} \phi} - \frac{\mathrm{d} \ln A}{\mathrm{d} \phi} T  \label{eqn:scalar-field-eqn-ef}
\end{align}
where $T = g^{\mu \nu} T_{\mu \nu}$ is the trace of the matter stress energy tensor defined as
\begin{equation}
T_{\mu \nu} \equiv \frac{-2}{\sqrt{-g}} \frac{\delta S_{\mathrm{mat}}}{\delta g^{\mu \nu}}\,.
\label{eqn:stress-energy-tensor-ef}
\end{equation}
To finish the stress-energy tensor of the scalar filed is
\begin{equation}\label{e.Tphi}
\begin{split}
T_{\mu \nu}^{(\phi)} &\equiv \frac{-2}{\sqrt{-g}} \frac{\delta S_{\phi}}{\delta g^{\mu \nu}} \\
&= \partial_{\mu} \phi \, \partial_{\nu} \phi - \frac{1}{2} g_{\mu \nu} g^{\rho \sigma} \partial_{\rho} \phi \, \partial_{\sigma} \phi - g_{\mu \nu} V(\phi) \, .
\end{split}
\end{equation}
It follows that the density and isotropic pressure measured by an observer comoving with $u^\mu$ (with {$u_\mu u^\mu=-1$}) are given by $\rho_\phi = T^{(\phi)} _{\mu\nu} u^\mu u^\nu$ and $3P_\phi-\rho_\phi = T^{(\phi)} _{\mu\nu} g^{\mu\nu}$.

Assuming a Minkowski spacetime in Einstein frame and a non-relativistic matter source,  the scalar field equation~(\ref{eqn:scalar-field-eqn-ef}) boils down to
\begin{equation}\label{eqn:scalar-field-eqn-newtonian-limit-ef}
-\ddot\phi+\Delta \phi = \frac{\mathrm{d} V}{\mathrm{d} \phi} + \frac{\mathrm{d} \ln A}{\mathrm{d} \phi} \rho \, ,
\end{equation}
and one has
\begin{eqnarray}
\rho_\phi ({\bf x})&=& \frac{1}{2}\dot\phi^2 + \frac{1}{2}\left(\nabla\phi\right)^2 + V \,, \label{eq.rhophidef}\\[5pt]
P_\phi ({\bf x})&=& \frac{1}{2}\dot\phi^2 - \frac{1}{6}\left(\nabla\phi\right)^2 - V \,, \label{eq.Pphidef}
\end{eqnarray}
with a dot refeering to $u^\mu\partial_\mu$. It is interesting to note that 
$$
\rho_\phi+3P_\phi= 2(\dot\phi^2 -V)
$$
hence sharing the same expression as for a homogeneous field, but still with $\phi({\bf x},t)$. Then,
$$
\rho_\phi+P_\phi= \dot\phi^2+\frac{1}{3}\left(\nabla\phi\right)^2 
$$
and the KG equation~(\ref{eqn:scalar-field-eqn-newtonian-limit-ef}) implies
$$
\dot \rho_\phi = –\alpha(\phi)\rho \dot\phi +\left(\dot\phi\Delta\phi +\nabla\phi.\nabla\dot\phi\right),
$$
with the first term describing the force of the standard matter on the scalar field fluid, {\em i.e.}, the counterpart of the fifth force, and the second term vanishing for a homogeneous field. It has to be compared with the similar equation one would get for a homogeneous scalar field in a FL spacetime
$$
\dot \rho_\phi +3H(\rho_\phi+P_\phi) = –\alpha(\phi)\rho \dot\phi.
$$

\section{General derivation of $\delta_{\rho}$ and $\delta_{p}$} \label{appE}

For a {\em static} configuration, it is clear from Eqs.~(\ref{eq.rhophidef}--\ref{eq.Pphidef}) that
\begin{eqnarray}
{\overline{\rho_{\phi}}}_{L_0}&=& \frac{1}{L_0^3}\int_\Omega\left[\frac{1}{2}\left(\nabla\phi\right)^2 + C\phi^p  \right] \dd^3{\bf x}\,,\nonumber \\
{\overline{P_{\phi}}}_{L_0}&=& \frac{1}{L_0^3}\int_\Omega\left[- \frac{1}{6}\left(\nabla\phi\right)^2 -  C\phi^p  \right]  \dd^3{\bf x} \nonumber
\end{eqnarray}
once the parametrization~(\ref{e.paraV}) for the potential has been used. The first term can be integrated by part and, once the KG equation is used to express the Laplacian, one gets
\begin{eqnarray}
{\overline{\rho_{\phi}}}_{L_0}&=& \frac{1}{L_0^3}\int_\Omega\left[ C\left(1-\frac{p}{2}\right)\phi^p -\frac{\beta}{2M_{\rm P}} \rho\phi\right] \dd^3{\bf x}\,, \label{e.Brho1}\\[5pt]
{\overline{P_{\phi}}}_{L_0}&=& \frac{1}{L_0^3}\int_\Omega\left[C\left(\frac{p}{6}-1\right)\phi^p +\frac{\beta}{6M_{\rm P}} \rho\phi\  \right]  \dd^3{\bf x}\,.\label{e.BP1}
\end{eqnarray}
Note that in the case of a non-dynamical field, but sill assuming a Minkowski background spacetime, the averaged energy density and pressure would take an additional contribution, respectively given by
$$
\frac{1}{L_0^3}\int_\Omega \frac{1}{2}\left[ \dot\phi^2- \phi\ddot\phi \right] \dd^3{\bf x}\,,\quad
\frac{1}{L_0^3}\int_\Omega \frac{1}{2}\left[ \dot\phi^2+\frac{1}{3}\phi\ddot\phi \right] \dd^3{\bf x}
$$
since Eq.~(\ref{e.kg1}) has en extra $-\ddot\phi$ term on its l.h.s., {\em i.e.}, using Eq.~(\ref{eqn:scalar-field-eqn-newtonian-limit-ef}).

Now, for a homogeneous distribution with density $\rho_0$ and reminding that $\rho_{\phi_{\rm macro}} = - P_{\phi_{\rm macro}} = V(\phi_{\rm macro}) = C\phi_{\rm macro}^p$, one gets
\begin{eqnarray}
\delta_\rho &=&  \left(\frac{\phi_0}{\phi_{\rm macro}}\right)^p \! \left[\left(1-\frac{p}{2} \right) \!\! \int_{\tilde\Omega}\!\! \tilde\phi^p\dd^3\tilde{\bf x}-\frac{|p|}{2}\!\! \int_{\tilde\Omega}\!\! \tilde\rho\tilde \phi\dd^3\tilde{\bf x}  \right]  \,, \label{e.Bdrho1}\\[5pt]
\delta_P &=& \left(\frac{\phi_0}{\phi_{\rm macro}}\right)^p\!\left[\left(1-\frac{p}{6} \right)\!\!  \int_{\tilde\Omega}\!\! \tilde\phi^p\dd^3\tilde{\bf x}-\frac{|p|}{6}\!\! \int_{\tilde\Omega} \! \!\tilde\rho\tilde \phi\dd^3\tilde{\bf x}  \right]  \,.\label{e.BdP1}
\end{eqnarray}
The prefactor can be rewritten in the different forms
$$
 \left(\frac{\phi_0}{\phi_{\rm max}}\right)^p= \left(\frac{\rho_0}{\rho_{\rm macro}}\right)^\frac{p}{p-1} =\left(\frac{4\pi}{3}\tilde R^3\right)^{-\frac{p}{p-1}},
$$
the last equality holding for homogeneous spheres. The homogeneous distribution limit is obtained when ${\mathrm{Vol}(\tilde{\Omega}) \equiv \frac{4\pi}{3}\tilde R^3 \rightarrow 1}$.

\section{Details on the numerical implementation} \label{appB}

The dimensionless Klein--Gordon equation~(\ref{e.KGred}) is solved using the \textit{femtoscope} software~\cite{Levy:2022xni} \ \textemdash \ a \textsc{Python} code\footnote{See \url{https://github.com/onera/femtoscope}.} that makes use of the Finite Element Method (FEM). The solutions of Eq.~(\ref{e.KGred}) are living in the Sobolev space of periodic functions $H^1_{\text{per}} (\tilde{\Omega})$, where $\tilde{\Omega}$ is the unit cube centered at the origin in $\mathbb{R}^3$. Denoting $H^1(\tilde{\Omega})$ the standard Sobolev space and given $\tilde{u} \in H^1(\tilde{\Omega})$, the periodic boundary conditions are written as
\begin{equation}
	\forall \begin{cases}
i \in \{ 1 , \, 2 , \, 3 \} \\
 \tilde{\mathbf{x}} \in \tilde{\Gamma} , \, \tilde{x}_i=0 
\end{cases} \hspace{-0.2cm} , \quad
\begin{cases}
\tilde{u}(\tilde{\mathbf{x}} + \tilde{\mathbf{e}}_i) = \tilde{u}(\tilde{\mathbf{x}}) \\
\tilde{\boldsymbol{\nabla}} \tilde{u}(\tilde{\mathbf{x}} + \tilde{\mathbf{e}}_i) = \tilde{\boldsymbol{\nabla}} \tilde{u}(\tilde{\mathbf{x}}) 
\end{cases} .
\label{eqn:epbc}
\end{equation}
Formally, one has
$$ H^1_{\text{per}}(\tilde{\Omega}) \coloneqq \Bigl\{ \tilde{u} \in H^1(\tilde{\Omega}) \mid \tilde{u} \text{ satisfies \eqref{eqn:epbc}}  \Bigr\} \, . $$
In finite element computations, conditions~(\ref{eqn:epbc}) are enforced by tying degrees of freedom belonging to opposite faces of the cube together. The actual implementation is done in the Sfepy~\cite{Sfepy} \textemdash \ the FEM software on top of which \textit{femtoscope} is built. Note that in the present case, of a ball centered in $\tilde{\Omega}$ (see Fig.~\ref{fig:geometry}), it is also possible to apply homogeneous Neumann boundary conditions on $\tilde{\Gamma}$ for symmetry reasons.

The sparse linear systems arising from the discretization of the relevant weak forms are solved with the MUMPS solver. In the case of the chameleon non-linear KG equation, we employ standard Newton's iterations with a relaxation parameter set to 0.4 to ensure convergence. The latter is also dependent on the choice of a sufficiently `good' initial guess. In this respect, we use $\tilde{\phi}_{\mathrm{macro}}$ for the unscreened regime, and approximation~(\ref{eqn:bivalued-phi}) in the screened regime.

\section{Details on the derivation of Eq.~(\ref{e.sphere1})} \label{appC}

In order to derive Eq.~(\ref{e.sphere1}), it is convenient to use a multipolar decomposition~\cite{abramowitz+stegun},
\begin{equation} \label{eq_as}
\frac{\rm{e}^{- \vert{\bf r}-\bf{s}\vert/\lambda}}{\vert{\bf r}-\bf{s}\vert} = \sum_{\ell=0}^{\infty}\frac{(2\ell+1)}{\sqrt{rs}}  K_{\ell+\frac{1}{2}}\left( \frac{r}{\lambda} \right) I_{\ell+\frac{1}{2}} \left( \frac{s}{\lambda} \right) P_\ell(\cos \varphi),\nonumber
\end{equation}
valid for  $r>s$, can be inserted in the general solution for the potential in terms of the Green function of the massive Klein--Gordon equation as
\begin{equation}
\phi({\bf r}) = -\frac{1}{4\pi}\frac{\beta}{M_{\rm P}} \int \rho({\bf s})\frac{\rm{e}^{-\vert{\bf r}-\bf{s}\vert/\lambda}}{\vert{\bf r}-\bf{s}\vert} \dd^3{\bf s}
\end{equation}
to derive the general solution of the field distribution. In these expressions, $P_\ell$ are Legendre polynomials, $I_{\ell+\frac{1}{2}}$ and $K_{n+\frac{1}{2}}$ are modified spherical Bessel functions of the second and third kinds. 

In the particular case of a homogeneous sphere of radius $R$ and density $\rho_0$, $ \dd^3{\bf s}=s^2\dd s\dd^2\Omega$ and the integration over the angle gives $4\pi\delta_{\ell,0}$ so that
\begin{equation}
\phi(r) =-\frac{\beta\rho_0}{M_{\rm P}}  \left\lbrace
\begin{tabular}{l}
$\int_0^R s^2 \dd s \frac{K_{1/2}(r/\lambda) I_{1/2}(s/\lambda)}{\sqrt{rs}}$   \\[8pt]
$\int_0^r s^2 \dd s \frac{K_{1/2}(r/\lambda) I_{1/2}(s/\lambda)}{\sqrt{rs}}$\\
  $\quad+ \int_r^R s^2 \dd s \frac{I_{1/2}(r/\lambda) K_{1/2}(s/\lambda)}{\sqrt{rs}}$\nonumber
\end{tabular}
\right.,
\end{equation}
the first line for $r>R$ and the second for $r<R$. Then, one needs to express the Bessel functions
$$
K_{1/2}(x)=\sqrt{\frac{\pi}{2}} \frac{\hbox{e}^{-x}}{\sqrt{x}}, \qquad
I_{1/2}(x)=\sqrt{\frac{2}{\pi}} \frac{\sinh x}{\sqrt{x}}
$$
to get
\begin{equation}
\phi(r) =-\frac{\beta\rho_0}{M_{\rm P}}\lambda  \left\lbrace
\begin{tabular}{l}
$\left[\int_0^R s \sinh\frac{s}{\lambda} \dd s\right] \frac{\hbox{e}^{-r/\lambda}}{r} $ \\ [8pt]
$\left[\int_0^r s \sinh\frac{s}{\lambda} \dd s\right] \frac{\hbox{e}^{-r/\lambda}}{r} $ \\ [5pt]
$\quad + \left[ \int_r^R\hbox{e}^{-s/\lambda} s \dd s  \right]\frac{\sinh (r/\lambda)}{r}$ 
\end{tabular}
\right.\nonumber
\end{equation}

It follows that
\begin{equation}
\phi(r) =-\frac{\beta\rho_0}{M_{\rm P}} \frac{R^3}{3}\left\lbrace
\begin{tabular}{l}
$\Phi(R/\lambda) \frac{\hbox{e}^{-r/\lambda}}{r} $\\[8pt]
$3\frac{1-\hbox{e}^{-R/\lambda}\frac{R+\lambda}{r}\sinh(r/\lambda) }{(R/\lambda)^3}\frac{1}{\lambda} $
\end{tabular}
\right.,
\end{equation}
with $\Phi$ defined in Eq.~(\ref{e.PPHI}). Equivalently, in terms of the total mass $M=4\pi \rho_0 R^3/3$, it reads
\begin{equation}
\phi(r) =-\frac{1}{4\pi}\frac{\beta M}{M_{\rm P}}  \left\lbrace
\begin{tabular}{l}
$\Phi(R/\lambda) \frac{\hbox{e}^{-r/\lambda}}{r} $\\[8pt]
$3\frac{1-\hbox{e}^{-R/\lambda}\frac{R+\lambda}{r}\sinh(r/\lambda) }{(R/\lambda)^3}\frac{1}{\lambda} $
\end{tabular}
\right..
\end{equation}
The solution is indeed regular in $r=R$ and matches the results by Ref.~\cite{Fleury:2016tsz}.

\section{Details on the derivation of Eq.~(\ref{e.delta_phi_analytical})}\label{appD}

In the so-called screened regime, the chameleon field takes the value that minimizes its effective potential everywhere in the ball except in a thin-shell of thickness $\epsilon \tilde{R}$ below its surface.\footnote{Following notations of Sec.~\ref{subsec3a}, we have $\epsilon = \Delta R / R$.} For convenience, let us denote $\mathcal{S} = \mathcal{B}[(1-\epsilon)\tilde{R}]$ the screened region where the field is constant, and $\mathrm{TS} = \mathcal{B}(\tilde{R}) \setminus \mathcal{S}$ the thin-shell region. Using the argument of \S~\ref{subsec3a}, one can easily show that
\begin{equation}
\int_{\tilde{\Omega} \setminus \mathcal{S}} \tilde{\phi}^{-(n+1)} \, \mathrm{d}^3 \tilde{\mathbf{x}} = \int_{\mathrm{TS}} \tilde{\rho} \, \mathrm{d}^3 \tilde{\mathbf{x}} \simeq 4 \pi \epsilon \tilde{R}^3 \, ,
\label{eqn:appD1}
\end{equation}
where the latter approximation is valid when $\epsilon \ll 1$. We can then use this to get an estimate of the value of the scalar field in the vacuum space surrounding the spherical particles. Indeed, if we assume that $\tilde{\phi}$ does not vary much in the interparticle space, we get from Eq.~(\ref{eqn:appD1}) that $\forall \tilde{\mathbf{x}} \in \tilde{\Omega} \setminus \mathcal{B}(\tilde{R})$,
\begin{equation}
\tilde{\phi} (\tilde{\mathbf{x}}) \simeq \tilde{\phi}_{\mathrm{vac}} = \left( \frac{4 \pi \epsilon \tilde{R}^3}{1 - 4 \pi \tilde{R}^3/3} \right)^{\! -1/(n+1)} .
\label{eqn:appD2}
\end{equation}
Note that in practice, the chameleon field is not strictly constant in vacuum. Therefore $\tilde{\phi}_{\mathrm{vac}}$ is a lower bound for the maximum value taken by the scalar field in $\tilde{\Omega}$.

However, for Eq.~(\ref{eqn:appD2}) to be useful is practice, one needs to determine a suitable value for the thin-shell parameter $\epsilon$, which is \textit{a priori} dependent on the dimensionless parameter $\tilde{\lambda}$. In the case of a single ball of density $\rho_{\mathrm{b}}$ immersed in a lower density background, Ref.~\cite{Khoury:2003rn} provides us with the following analytical approximation
$$ \epsilon \simeq \frac{\phi_{\mathrm{vac}} - \phi_{\mathrm{b}}}{6 \beta M_{\rm P} \Phi_{\mathrm{b}}} \, , $$
where $\Phi_{\mathrm{b}} = M_{\mathrm{b}} / (8 \pi M_{\rm P}^2 R)$ is the Newtonian potential at the surface of the homogeneous sphere of mass $M_{\mathrm{b}}$ and radius $R$. Re-writing the above equation in terms of dimensionless quantities yields
\begin{equation}
\epsilon \simeq ( \tilde{\lambda} / \tilde{R} )^{\! 2} \bigl( \tilde{\phi}_{\mathrm{vac}} - 1 \bigr) \, .
\label{eqn:appD3}
\end{equation}
Plugging back Eq.~(\ref{eqn:appD3}) into Eq.~(\ref{eqn:appD2}) yields $\tilde{\phi}_{\mathrm{vac}}$ as given by Eq.~(\ref{e.phi_vac_analytical}).

Finally, we obtain the analytical approximation of $\delta_{\phi}$ given by Eq.~(\ref{e.delta_phi_analytical}) by approximating $\tilde{\phi}$ as a bi-valued function, where
\begin{equation}
\tilde{\phi} =
\begin{cases}
1 & \text{in the screened region } \mathcal{S \, ,} \\[3pt]
\tilde{\phi}_{\mathrm{vac}} & \text{in } \tilde{\Omega} \setminus \mathcal{S} \, .
\end{cases}
\label{eqn:bivalued-phi}
\end{equation}

\bibliography{BIBHO}

\begin{thebibliography}{10}

\bibitem{Ellis:1971pg}
G.~F.~R.~Ellis
``Relativistic cosmology''
in {\it General Relativity and Cosmology}, Enrico Fermi Summer School Course XLVII ed R.~K. Sachs
(New York:Academic, 1971) p.~104-182.

\bibitem{Ellis:1984bqf}
G.~F.~R.~Ellis,
``Relativistic Cosmology: Its Nature, Aims and Problems'',
Fundam. Theor. Phys. \textbf{9}, 215-288 (1984)

\bibitem{Ellis:1987zz}
G.~F.~R.~Ellis and W.~Stoeger,
``The 'fitting problem' in cosmology'',
Class. Quant. Grav. \textbf{4}, 1697-1729 (1987)

\bibitem{Ellis:2005uz}
G.~F.~R.~Ellis and T.~Buchert,
``The Universe seen at different scales'',
Phys. Lett. A \textbf{347}, 38-46 (2005)
[arXiv:gr-qc/0506106 [gr-qc]].

\bibitem{Clarkson:2011zq}
C.~Clarkson, G.~Ellis, J.~Larena and O.~Umeh,
``Does the growth of structure affect our dynamical models of the universe? The averaging, backreaction and fitting problems in cosmology'',
Rept. Prog. Phys. \textbf{74}, 112901 (2011)
[arXiv:1109.2314 [astro-ph.CO]].

\bibitem{Buchert:1999er}
T.~Buchert,
``On average properties of inhomogeneous fluids in general relativity. 1. Dust cosmologies'',
Gen. Rel. Grav. \textbf{32}, 105-125 (2000)
[arXiv:gr-qc/9906015 [gr-qc]].

\bibitem{Buchert:2001sa}
T.~Buchert,
``On average properties of inhomogeneous fluids in general relativity: Perfect fluid cosmologies'',
Gen. Rel. Grav. \textbf{33}, 1381-1405 (2001)
[arXiv:gr-qc/0102049 [gr-qc]].

\bibitem{Zalaletdinov:1996aj}
R.~M.~Zalaletdinov,
``Averaging problem in general relativity, macroscopic gravity and using Einstein's equations in cosmology'',
Bull. Astron. Soc. India \textbf{25}, 401-416 (1997)
[arXiv:gr-qc/9703016 [gr-qc]].

\bibitem{Zalaletdinov:2008ts}
R.~Zalaletdinov,
``The Averaging Problem in Cosmology and Macroscopic Gravity'',
Int. J. Mod. Phys. A \textbf{23}, 1173-1181 (2008)
[arXiv:0801.3256 [gr-qc]].

\bibitem{Paranjape:2008ai}
A.~Paranjape and T.~P.~Singh,
``Structure Formation, Backreaction and Weak Gravitational Fields'',
JCAP \textbf{03}, 023 (2008)
[arXiv:0801.1546 [astro-ph]].

\bibitem{Buchert:2002ht}
T.~Buchert and M.~Carfora,
``Regional averaging and scaling in relativistic cosmology'',
Class. Quant. Grav. \textbf{19}, 6109-6145 (2002)
[arXiv:gr-qc/0210037 [gr-qc]].

\bibitem{Calzetta:1999zr}
E.~A.~Calzetta, B.~L.~Hu and F.~D.~Mazzitelli,
``Coarse grained effective action and renormalization group theory in semiclassical gravity and cosmology'',
Phys. Rept. \textbf{352}, 459-520 (2001)
[arXiv:hep-th/0102199 [hep-th]].

\bibitem{Carfora:1995fj}
M.~Carfora and K.~Piotrkowska,
``A Renormalization group approach to relativistic cosmology'',
Phys. Rev. D \textbf{52}, 4393-4424 (1995)
[arXiv:gr-qc/9502021 [gr-qc]].

\bibitem{Futamase:1996fk}
T.~Futamase,
``Averaging of a locally inhomogeneous realistic universe'',
Phys. Rev. D \textbf{53}, 681-689 (1996)

\bibitem{Boersma:1997yt}
J.~P.~Boersma,
``Averaging in cosmology'',
Phys. Rev. D \textbf{57}, 798-810 (1998)
[arXiv:gr-qc/9711057 [gr-qc]].

\bibitem{Stoeger:1999ig}
W.~R.~Stoeger, S.J., A.~Helmi and D.~F.~Torres,
``Averaging Einstein's equations: The Linearized case'',
Int. J. Mod. Phys. D \textbf{16}, 1001-1026 (2007)
[arXiv:gr-qc/9904020 [gr-qc]].

\bibitem{Fleury:2013sna}
P.~Fleury, H.~Dupuy and J.~P.~Uzan,
``Interpretation of the Hubble diagram in a nonhomogeneous universe'',
Phys. Rev. D \textbf{87}, no.12, 123526 (2013)
[arXiv:1302.5308 [astro-ph.CO]].

\bibitem{Fleury:2013uqa}
P.~Fleury, H.~Dupuy and J.~P.~Uzan,
``Can all cosmological observations be accurately interpreted with a unique geometry?'',
Phys. Rev. Lett. \textbf{111}, 091302 (2013)
[arXiv:1304.7791 [astro-ph.CO]].

\bibitem{Ben-Dayan:2012ccq}
I.~Ben-Dayan, M.~Gasperini, G.~Marozzi, F.~Nugier and G.~Veneziano,
``Do stochastic inhomogeneities affect dark-energy precision measurements?'',
Phys. Rev. Lett. \textbf{110}, no.2, 021301 (2013)
[arXiv:1207.1286 [astro-ph.CO]].

\bibitem{RWrefs}
Y. B. Zel’dovich, Sov. Astron. Lett. {\bf 8}, 13 (1964); 
V. M. Dashevskii and Y. B. Zel’dovich, Sov. Astronom. {\bf 8}, 854 (1965);
 J. E. Gunn, Astrophys. J. {\bf 150}, 737 (1967); 
 R. Kantowski, Astrophys. J. {\bf 155}, 89 (1969); 
 C. Dyer and R. Roeder, Astrophys. J. {\bf 174}, L115 (1972); 
 S. Weinberg, Astrophys. J. Lett. {\bf 208}, L1 (1976); 
 D. E. Holz and R. M. Wald, Phys. Rev. D {\bf 58}, 063501 (1998);  
 K. Bolejko and P. G. Ferreira, JCAP {\bf 5}, 003 (2012).

\bibitem{Clarkson:2011br}
C.~Clarkson, G.~F.~R.~Ellis, A.~Faltenbacher, R.~Maartens, O.~Umeh and J.~P.~Uzan,
``(Mis-)Interpreting supernovae observations in a lumpy universe'',
Mon. Not. Roy. Astron. Soc. \textbf{426}, 1121-1136 (2012)
[arXiv:1109.2484 [astro-ph.CO]].

\bibitem{Fleury:2015rwa}
P.~Fleury, J.~Larena and J.~P.~Uzan,
``The theory of stochastic cosmological lensing'',
JCAP \textbf{11}, 022 (2015)
[arXiv:1508.07903 [gr-qc]].

\bibitem{Fleury:2017owg}
P.~Fleury, J.~Larena and J.~P.~Uzan,
``Weak gravitational lensing of finite beams'',
Phys. Rev. Lett. \textbf{119}, no.19, 191101 (2017)
[arXiv:1706.09383 [gr-qc]].

\bibitem{Fleury:2018cro}
P.~Fleury, J.~Larena and J.~P.~Uzan,
``Cosmic convergence and shear with extended sources'',
Phys. Rev. D \textbf{99}, no.2, 023525 (2019)
[arXiv:1809.03919 [astro-ph.CO]].

\bibitem{Ferreira:1997au}
P.~G.~Ferreira and M.~Joyce,
``Structure formation with a selftuning scalar field'',
Phys. Rev. Lett. \textbf{79}, 4740-4743 (1997)
[arXiv:astro-ph/9707286 [astro-ph]].

\bibitem{Creminelli:2008wc}
P.~Creminelli, G.~D'Amico, J.~Norena and F.~Vernizzi,
``The Effective Theory of Quintessence: the w\ensuremath{<}-1 Side Unveiled'',
JCAP \textbf{02}, 018 (2009)
[arXiv:0811.0827 [astro-ph]].

\bibitem{Creminelli:2009mu}
P.~Creminelli, G.~D'Amico, J.~Norena, L.~Senatore and F.~Vernizzi,
``Spherical collapse in quintessence models with zero speed of sound'',
JCAP \textbf{03}, 027 (2010)
[arXiv:0911.2701 [astro-ph.CO]].

\bibitem{Uzan:1999ch}
J.~P.~Uzan,
``Cosmological scaling solutions of nonminimally coupled scalar fields'',
Phys. Rev. D \textbf{59}, 123510 (1999)
[arXiv:gr-qc/9903004 [gr-qc]].

\bibitem{Levy:2024vyd}
H.~L\'evy and J.~P.~Uzan,
``Testing screened scalar-tensor theories of gravity with atomic clocks'',
Phys. Rev. D \textbf{111}, no.6, 064012 (2025)
[arXiv:2410.17292 [gr-qc]].

\bibitem{Brax:2004qh}
P.~Brax, C.~van de Bruck, A.~C.~Davis, J.~Khoury and A.~Weltman,
``Detecting dark energy in orbit: The cosmological chameleon'',
Phys. Rev. D \textbf{70}, 123518 (2004)
[arXiv:astro-ph/0408415 [astro-ph]].

\bibitem{Brax:2004px}
P.~Brax, C.~van de Bruck, A.~C.~Davis, J.~Khoury and A.~Weltman,
``Chameleon dark energy'',
AIP Conf. Proc. \textbf{736}, no.1, 105-110 (2004)
[arXiv:astro-ph/0410103 [astro-ph]].


\bibitem{Mota:2006fz}
D.~F.~Mota and D.~J.~Shaw,
``Evading Equivalence Principle Violations, Cosmological and other Experimental Constraints in Scalar Field Theories with a Strong Coupling to Matter,''
Phys. Rev. D \textbf{75} (2007), 063501
[arXiv:hep-ph/0608078 [hep-ph]].

\bibitem{Mota:2006ed}
D.~F.~Mota and D.~J.~Shaw,
``Strongly coupled chameleon fields: New horizons in scalar field theory,''
Phys. Rev. Lett. \textbf{97} (2006), 151102
[arXiv:hep-ph/0606204 [hep-ph]].


\bibitem{Levy:2022xni}
H.~L\'evy, J.~Berg\'e and J.~P.~Uzan,
``Solving nonlinear Klein--Gordon equations on unbounded domains via the finite element method'',
Phys. Rev. D \textbf{106}, no.12, 124021 (2022)
[arXiv:2209.07226 [gr-qc]].

\bibitem{Levy:2024mut}
H.~L\'evy,
``Towards well-posed and versatile numerical solutions of scalar-tensor theories of gravity with screening mechanisms : applications at sub-Solar system scales'',
tel-04789073 (PhD thesis, 2024).

\bibitem{Levy:2023tps}
H.~L\'evy, J.~Berg\'e and J.~P.~Uzan,
``What to expect from scalar-tensor space geodesy'',
Phys. Rev. D \textbf{109}, no.8, 084009 (2024)
[arXiv:2310.03769 [gr-qc]].

\bibitem{Khoury:2003rn}
J.~Khoury and A.~Weltman,
``Chameleon cosmology'',
Phys. Rev. D \textbf{69}, 044026 (2004)
[arXiv:astro-ph/0309411 [astro-ph]].

\bibitem{Khoury:2003aq}
J.~Khoury and A.~Weltman,
``Chameleon fields: Awaiting surprises for tests of gravity in space'',
Phys. Rev. Lett. \textbf{93}, 171104 (2004)
[arXiv:astro-ph/0309300 [astro-ph]].

\bibitem{Hinterbichler:2010es}
K.~Hinterbichler and J.~Khoury,
``Symmetron Fields: Screening Long-Range Forces Through Local Symmetry Restoration'',
Phys. Rev. Lett. \textbf{104}, 231301 (2010)
[arXiv:1001.4525 [hep-th]].

\bibitem{Sanghai:2015wia}
V.~A.~A.~Sanghai and T.~Clifton,
``Post-Newtonian Cosmological Modelling'',
Phys. Rev. D \textbf{91}, 103532 (2015)
[erratum: Phys. Rev. D \textbf{93}, no.8, 089903 (2016)]
[arXiv:1503.08747 [gr-qc]].

\bibitem{Sanghai:2016ucv}
V.~A.~A.~Sanghai and T.~Clifton,
``Cosmological backreaction in the presence of radiation and a cosmological constant'',
Phys. Rev. D \textbf{94}, no.2, 023505 (2016)
[arXiv:1604.06345 [gr-qc]].

\bibitem{Sanghai:2017yyn}
V.~A.~A.~Sanghai, P.~Fleury and T.~Clifton,
``Ray tracing and Hubble diagrams in post-Newtonian cosmology'',
JCAP \textbf{07}, 028 (2017)
[arXiv:1705.02328 [astro-ph.CO]].


\bibitem{Sikora:2018imr}
S.~Sikora and K.~G\l{}\'od,
``Perturbatively constructed cosmological model with periodically distributed dust inhomogeneities'',
Phys. Rev. D \textbf{99}, no.8, 083521 (2019)
[arXiv:1811.06836 [gr-qc]].

\bibitem{Briddon:2024ftz}
C.~Briddon, T.~Clifton and P.~Fleury,
``Emergent cosmological expansion in scalar\textendash{}tensor theories of gravity'',
Class. Quant. Grav. \textbf{42}, no.1, 015013 (2025)
[arXiv:2406.01397 [gr-qc]].

\bibitem{Fleury:2016tsz}
P.~Fleury,
``Cosmic backreaction and Gauss\textquoteright{}s law'',
Phys. Rev. D \textbf{95}, no.12, 124009 (2017)
[arXiv:1609.03724 [gr-qc]].

\bibitem{will}
 C.M. Will, {\it Theory and Experiment in Gravitational Physics} (Cambridge University Press, Cambridge; New York, 1993).

\bibitem{blanchet} 
L. Blanchet, ``Gravitational radiation from post-Newtonian sources and inspiralling compact binaries", Living Rev. Relativ {\bf9}:4 (2006) [https://doi.org/10.12942/lrr-2006-4. grqc/
0202016].


\bibitem{Pitrou:2023swx}
C.~Pitrou and J.~P.~Uzan,
``Hubble Tension as a Window on the Gravitation of the Dark Matter Sector'',
Phys. Rev. Lett. \textbf{132}, no.19, 191001 (2024)
[arXiv:2312.12493 [astro-ph.CO]].

\bibitem{Uzan:2023dsk}
J.~P.~Uzan and C.~Pitrou,
``Hubble tension as a window on the gravitation of the dark matter sector: Exploration of a family of models'',
Phys. Rev. D \textbf{109}, no.10, 103505 (2024)
[arXiv:2312.12408 [astro-ph.CO]].

\bibitem{Esposito-Farese:2000pbo}
G.~Esposito-Farese and D.~Polarski,
``Scalar tensor gravity in an accelerating universe'',
Phys. Rev. D \textbf{63}, 063504 (2001)
[arXiv:gr-qc/0009034 [gr-qc]].

\bibitem{Uzan:2006mf}
J.~P.~Uzan,
``The acceleration of the universe and the physics behind it'',
Gen. Rel. Grav. \textbf{39}, 307-342 (2007)
[arXiv:astro-ph/0605313 [astro-ph]].

\bibitem{Uzan:2024ded}
J.~P.~Uzan,
``Fundamental constants: from measurement to the universe, a window on gravitation and cosmology'',
[arXiv:2410.07281 [astro-ph.CO]].

\bibitem{Burrage:2017qrf}
C.~Burrage and J.~Sakstein,
``Tests of Chameleon Gravity'',
Living Rev. Rel. \textbf{21}, no.1, 1 (2018)
[arXiv:1709.09071 [astro-ph.CO]].

\bibitem{Pernot-Borras:2019gqs}
M.~Pernot-Borr\`as, J.~Berg\'e, P.~Brax and J.~P.~Uzan,
``General study of chameleon fifth force in gravity space experiments",
Phys. Rev. D \textbf{100}, no.8, 084006 (2019)
[arXiv:1907.10546 [gr-qc]].

\bibitem{Pernot-Borras:2020jev}
M.~Pernot-Borr\`as, J.~Berg\'e, P.~Brax and J.~P.~Uzan,
``Fifth force induced by a chameleon field on nested cylinders'',
Phys. Rev. D \textbf{101}, no.12, 124056 (2020)
[arXiv:2004.08403 [gr-qc]].

\bibitem{Pernot-Borras:2021edr}
M.~Pernot-Borr\`as, J.~Berg\'e, P.~Brax, J.~P.~Uzan, G.~M\'etris, M.~Rodrigues and P.~Touboul,
``Constraints on chameleon gravity from the measurement of the electrostatic stiffness of the MICROSCOPE mission accelerometers'',
Phys. Rev. D \textbf{103}, no.6, 064070 (2021)
[arXiv:2102.00023 [gr-qc]].

\bibitem{Sfepy}
C.~Robert,
``Fast evaluation of finite element weak forms using python tensor contraction packages'',
Advances in Engineering Software 159, 103033 (2021)

\bibitem{abramowitz+stegun} 
M. Abramowitz, and I.A.  Stegun, 
Handbook of Mathematical Functions with Formulas, Graphs, and  Mathematical ables, (Dover Publications, New York, 1964).

\end{thebibliography}
\end{document}